\documentclass[aps,twocolumn, prb]{revtex4-1}	
\usepackage{graphicx}
\usepackage{amsmath, amsfonts,amssymb}
\usepackage{color}

\newcommand{\be}{\begin{equation}}
\newcommand{\ee}{\end{equation}}
\newcommand{\bee}{\begin{equation*}}
\newcommand{\eee}{\end{equation*}}
\newcommand{\ba}{\begin{align*}}
\newcommand{\ea}{\end{align*}}

\newcommand{\Tr}{\textrm{Tr}}
\newcommand{\Pf}{\textrm{Pf}}

\newcommand{\kv}{\mathbf{k}}

\newcommand{\tauv}{\vec{\tau}}

\newcommand{\dt}{\partial_t}
\newcommand{\dkx}{\partial_{k_x}}
\newcommand{\dky}{\partial_{k_y}}

\newcommand{\sgn}{\textrm{sgn}}
\newcommand{\nedge}{n_{\textrm{edge}}}

\newcommand{\Heff}{H_\textrm{eff}}

\newcommand{\bea}{\begin{eqnarray}}
\newcommand{\eea}{\end{eqnarray}}

\renewcommand{\vec}[1]{{\bf #1}}
\renewcommand{\epsilon}{\varepsilon}
\newcommand{\kinv}{\kv_\textrm{inv}}

\begin{document}

\title{Topological singularities and the general classification of Floquet-Bloch systems} 
\author{Frederik Nathan and Mark S. Rudner}
\affiliation{Niels Bohr International Academy and Center for Quantum Devices, University of Copenhagen, 2100 Copenhagen, Denmark}
\begin{abstract}
Recent works have demonstrated that the Floquet-Bloch bands of periodically-driven systems feature a richer topological structure than their non-driven counterparts.
The additional structure in the driven case arises from the periodicity of quasienergy, the energy-like quantity that defines the spectrum of a periodically-driven system.
Here we develop a new paradigm for the topological classification of Floquet-Bloch bands, based on the time-dependent spectrum of the driven system's evolution operator throughout one driving period.
Specifically, we show that this spectrum may host topologically-protected degeneracies at intermediate times, which control the topology of the Floquet bands of the full driving cycle.
This approach provides a natural framework for incorporating the role of symmetries, enabling a unified and complete classification of Floquet-Bloch bands and yielding new insight into the topological features that distinguish driven and non-driven systems.

\end{abstract} 

\maketitle

After the discovery\cite{Klitzing, TsuiFQH} and explanation\cite{TKNN, LaughlinQuantization, Halperin, AvronPRL1983, LaughlinFQH} of the quantized Hall effects, topology gained new importance as a mechanism for generating extremely robust quantum mechanical phenomena.
The realization that the Bloch bands of solid state systems could possess non-trivial topological characteristics led to the prediction\cite{Kane_Mele_1, Bernevig_Prediction, Fu_and_Kane_Prediction} and experimental discovery\cite{MolenkampQSH, Hasan3DTI}  of whole new classes of materials\cite{HasanKaneRMP, Bernevig} -- the topological insulators and superconductors -- which host a variety of remarkable and potentially useful phenomena. 
On a theoretical level, a complete topological classification\cite{Periodic_Table_2, Periodic_Table_1} of such systems has been developed, predicting a number of new phases.
However, finding materials that realize these phases remains a very challenging task, with no known examples for many topological classes. 

Motivated by the great successes and open challenges in the arena of topological matter, 
many authors have begun to explore the possibilities for realizing topological phenomena in {\it driven} quantum systems\cite{Niu2007,Photovoltaic_graphene, KBRD, Inoue2010, Lindner2011, 1D_Majorana, Lindner2013, Gu11, Floquet_Transport_Kitagawa, Delplace2013, Podolsky2013, Chiral_1D, TorresPRL2014, ErhaiPRL, Iadecola2013, Goldman2014, Dehghani2014, Grushin2014, Titum2015, Bilitewski2015, Seetharam2015}.
Time-dependent driving offers the opportunity to control a material's properties in a variety of new ways, potentially opening new routes for studying topological phenomena in solid state\cite{Wang2013}, atomic\cite{1D_Majorana, Jotzu2014, JimenezGarcia2014}, and optical systems\cite{Observation_in_Photonic_Qws, Rechtsman2013}.

Intriguingly, 
driven systems may host an even richer array of topological phenomena than their non-driven counterparts.
To date several examples of topological phenomena which can {\it only} be realized in driven systems have been found\cite{KBRD,1D_Majorana, Chiral_1D, Winding_Number, 2D_TR}, such as the existence of robust chiral edge states in two dimensional systems whose Floquet bands have trivial Chern indices\cite{Winding_Number}, and pairs of non-degenerate Majorana end modes with protected quasienergy splittings in one-dimensional systems\cite{1D_Majorana}.
This indicates that periodically driven systems 
feature additional topological structure beyond that found in non-driven systems.
However, a unifying 
 principle for understanding and classifying these new phenomena remains lacking.


\begin{figure}\begin{center}
\includegraphics[scale=0.58]{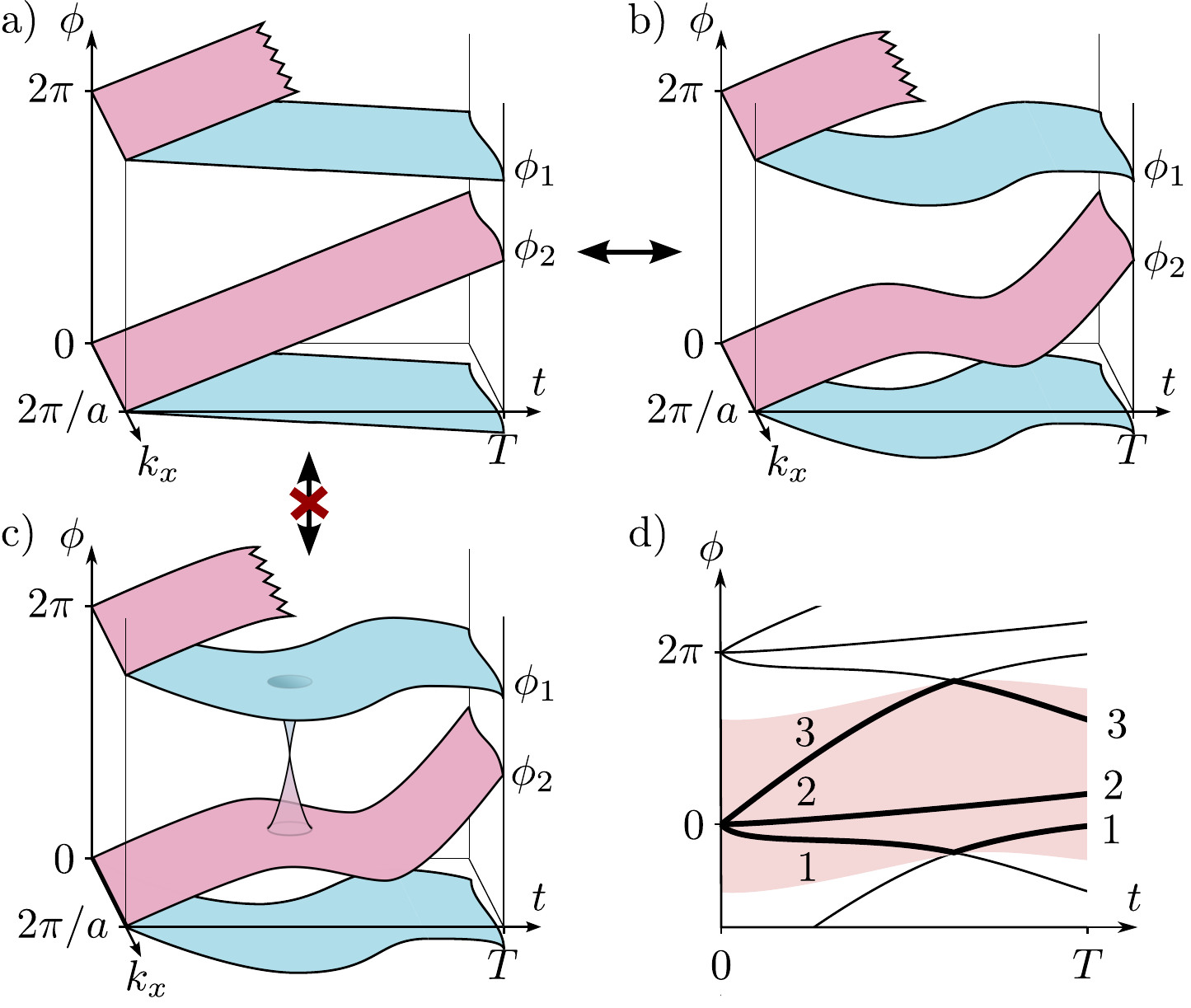}
\caption{Phase band representation of the evolution operator $U(\vec{k}, t)$, Eq.~(\ref{eq:SpectralDecomp}).
a) For a non-driven system, the phase eigenvalues grow linearly in time.
b) Here we show phase bands of a periodically driven system which are non-degenerate for all $\vec{k}$ and $t$. 
The evolution can be smoothly deformed into one 
obtainable in a non-driven system without closing any quasienergy gaps.
c) In this case the evolution operator features non-removable degeneracies which prevent such a deformation. 
The evolution is therefore topologically distinct from that of any non-driven system. 
d) Illustration of the phase band labeling scheme defined in Sec.~\ref{sec:PhaseBands}.
The shaded region indicates the ``phase Brillouin zone.''
}
%
\label{fig: Deformation 1}
\end{center}
\vspace{-0.2 in}
\end{figure}

In this work we answer the question: under what conditions does the evolution of a driven system become topologically distinct from that of a non-driven system? 
In doing so we develop a powerful and general framework that can be used to understand the topology of periodically driven systems.

In the analysis of periodically driven systems, the {\it Floquet operator}, denoted $U(T)$, plays a central role 
as the stroboscopic evolution operator that propagates the system forward in time through each complete driving period, $T$.
The spectrum of the Floquet operator, given by $U(T)|\Psi_n\rangle=e^{-i\varepsilon_n T}|\Psi_n\rangle$, plays an analogous role to the spectrum of the Hamiltonian in a non-driven system, with real-valued energies replaced by periodically-defined {\it quasienergies}, $\epsilon_n + 2\pi N/T = \epsilon_n$ for any integer $N$. 
For a system on a lattice, the single particle spectrum forms bands, the so-called Floquet bands.
Throughout this work we focus on systems defined on a lattice, with a finite number of bands.
While knowledge of the Floquet bands is sufficient to understand many aspects of the dynamics of a driven system, it was recently shown that the topological properties of the evolution are in particular {\it not} described by $U(T)$ alone\cite{Winding_Number}.
A proper description of the topology of driven systems must take into account the full evolution $U(t)$ for times $t$ throughout the entire driving period, $0 \le t \le T$.


As a means of elucidating the nature of the evolution $U(t)$, we focus on the ``phase bands'' of the system, i.e., 
the time-dependent spectrum 
of the system's evolution operator 
throughout one driving period. 
As depicted 
in Fig.~\ref{fig: Deformation 1}, for each time $t$ in the interval $0 \le t \le T$, the eigenvalues $\{e^{i\phi(\vec{k},t)}\}$ of the Bloch evolution operator $U(\vec{k}, t)$ form bands as a function of the crystal momentum $\vec{k}$.
For illustration we 
use a ``repeated zone'' representation for the phase bands, though the complete spectrum is contained within a single phase Brillouin zone of width $2\pi$, as indicated by the shaded region in Fig.~\ref{fig: Deformation 1}d. 
As a function of time, 
these phase bands 
form sheets which, along with the corresponding eigenvectors, contain full information about the evolution of the system.

Below we examine smooth deformations of the phase bands which, keeping the Floquet operator $U(T)$ fixed, determine when a given system's evolution can be smoothly deformed into one obtainable in a non-driven system. 
At time $t = 0$ the evolution is the identity.
Therefore all phase bands must originate with phases $\phi$ which are integer multiples of $2\pi$.
For a non-driven system with Hamiltonian $H$, the evolution operator is given by $U(t) = e^{-iHt}$.  
In this case the phase bands diverge from one another {\it linearly} in time due to the linear phase winding $\phi = Et$ for each eigenstate of $H$ with energy $E$ (see Fig.~\ref{fig: Deformation 1}a).
For the case of a driven system as shown in Fig.~\ref{fig: Deformation 1}b, the phase bands can be straightened through a continuous deformation, such that the evolution becomes indistinguishable from one generated by a time-independent Hamiltonian.
Crucially, as we show below, phase bands may be connected via topologically-protected degeneracies, or ``topological singularities,'' which {\it prevent} the evolution from being deformed into the canonical form for 
a non-driven system (Fig.~\ref{fig: Deformation 1}c).
These singularities  play a central role in defining the topology of periodically driven systems.

After establishing the existence of topological singularities in the bulk evolution, we study their ramifications for the edge mode spectrum of $U(T)$ for a two-dimensional system defined in a 
geometry with edges.
To this end we identify a complete set of independent topological quantities which are (by definition) invariant under any smooth deformation of the bulk evolution that preserves the Floquet operator $U(\vec{k},T)$.
Noting that the net number of chiral edge modes in each gap must also be invariant under such smooth deformations, we derive a bulk-edge correspondence in terms of these invariant quantities [Eq.~(\ref{eq: Winding number sum}) below].
We then show how the method can be extended to systems in arbitrary dimensions, also including the role of symmetries, thus providing means for a complete topological classification of Floquet-Bloch systems.
We show that symmetries considered previously, e.g.~in Refs.~\onlinecite{1D_Majorana, 2D_TR}, which generalize the Altland-Zirnbauer symmetry classes to the case of periodically driven systems, can be naturally incorporated into the phase-band picture. 
Importantly, we find that these symmetries can protect new types of topological singularities in the bulk.

In general we find that, for each bulk gap, the edge mode spectrum of a driven system in a given symmetry class has the same set of protected features as that of a non-driven system in the corresponding class.
However, the {\it global} edge mode spectrum and the relation between edge modes and bulk bands 
can be quite different.
Examples of such new or ``anomalous'' edge phenomena include Floquet-Majorana edge modes\cite{1D_Majorana, Kundu2013} with quasi-energy $\pi/T$ and chiral edge modes\cite{Winding_Number} in a 2D system with topologically trivial bulk Floquet bands. 
Here we also show that 
periodic driving, for example, allows  \textit{two-band} systems with time-reversal symmetry to have helical edge modes, while a minimum of four bands is required in the non-driven case.

Interestingly, we find that all the above phenomena are closely connected with the appearance of topological singularities in the bulk evolution.
Due to the additional freedom presented by time-dependence, we further speculate that there may be other new types of symmetry conditions (beyond those familiar from non-driven systems) which can protect new types of topological singularities and anomalous edge mode phenomena.

The remainder of the paper is structured as follows.
In Sec.~\ref{sec:PhaseBands} we formalize the description of phase bands, and characterize the singularities which may prevent them from being deformed into a trivial configuration.
Then in Sec.~\ref{sec:2dCase} we cast the topological characterization of two-dimensional systems (without symmetries) in terms of the phase bands and their singularities, giving new insight into the winding number invariants found previously in Ref.~\onlinecite{Winding_Number}.
In Sec.~\ref{cha:SYM} we show how additional symmetries (e.g., time reversal or particle-hole symmetry) can be naturally incorporated into this picture through their abilities to protect new types of singularities. 
Finally, in Sec.~\ref{sec:Discussion} we summarize our results and discuss the outlook for future work.
Technical aspects of derivations are provided in appendices.
\section{Phase bands of the evolution operator}
\label{sec:PhaseBands}
We now study the question of when the evolution of a periodically-driven Floquet-Bloch system can be considered topologically distinct from that of a non-driven system. 
In order to do this, we begin by defining 
the phase band picture of Floquet-Bloch evolution. 
In this section we focus on ``bulk'' systems with discrete translation symmetry 
(with infinite extent or periodic boundary conditions).
Here, the crystal momentum $\vec{k}$ is a good quantum number.
For now we leave the number of spatial dimensions arbitrary.

The evolution of a periodically-driven quantum system may equivalently be prescribed in terms of either a Hamiltonian $H(t + T) = H(t)$, where $T$ is the driving period, or by the corresponding evolution operator $U(t) = \mathcal{T}e^{-i\int_0^t H(t')dt'}$, where $\mathcal{T}$ denotes time ordering.
In this paper we primarily work directly with the evolution operator $U(t)$, which most clearly exposes the topological features of the evolution.
Importantly, although the Hamiltonian satisfies $H(t + T) = H(t)$, the evolution operator $U(t)$ is generally {\it not} periodic in time. 

For bulk systems,  crystal momentum $\vec{k}$ and time $t$ parametrize a family of Bloch evolution operators $U(\vec{k},t)$, which act within the space of periodic Bloch functions. 
When the time-dependent Hamiltonian is local and bounded, 
$U(\vec{k},t)$ is continuous in crystal momentum and time.



As an important first step in our analysis, we express $U(\vec{k},t)$ in terms of its spectral decomposition 
\be 
\label{eq:SpectralDecomp}U(\kv,t)=\sum_{n=1}^N P_n(\kv,t) e^{-i\phi_n(\kv,t)},
\ee
where $P_n(\kv,t)$ is the projector onto the $n$-th eigenstate  of $U(\kv,t)$ and $e^{-i\phi_n(\kv,t)}$ is the corresponding eigenvalue.
Here $N$ is the number of bands in the system.

We  refer to the functions $\{\phi_n(\kv,t)\}$ as the \textit{phase bands} of the system. 
In contrast to the 
quasienergy bands associated with a driven system's Floquet operator $U(\vec{k}, T)$, the phase bands 
depend  on time, and are continuously defined {\it throughout an entire driving cycle}, $0 \le t \le T$. 
At 
time $t=T$, the phase bands 
coincide with the system's Floquet bands. 
An illustration of phase bands for a one-dimensional system with two bands is shown in Fig.~\ref{fig: Deformation 1}. 

To resolve the ambiguity of the labeling of eigenstates of $U(\vec{k},t)$ we now define a prescription for assigning the values of the $n$ indices.
We focus on the phase bands $\{\phi_n(\vec{k},t)\}$, and work in a repeated zone representation where the spectrum is copied and shifted through all integer multiples of $2\pi$.
Recalling that $U(\vec{k}, 0) = \bf{1}$, each phase eigenvalue 
must start from an integer multiple of $2\pi$. 
However, as mentioned above, the full spectrum $\{e^{-i\phi_n(\vec{k},t)}\}$ of $U(\vec{k},t)$ is contained within one ``phase Brillouin zone.'' 

While in principle we could choose $\phi_n(\vec{k},0)$ to be equal to any integer multiple of $2\pi$, we choose to work in a fundamental phase Brillouin zone in which all phases originate from zero, i.e., $\phi_n(\vec{k},0) = 0$, for $n = 1\ldots N$ (see bold curves in Fig.~\ref{fig: Deformation 1}d). 
Next, we demand that each $\phi_n(\vec{k}, t)$ is a real-valued continuous function\cite{footnote:continuous_phases} 
of both $\vec{k}$ and $t$. 
Finally, we impose an ordering condition: if $\phi_n(\kv,t)\geq\phi_m(\kv,t)$ for one point in $\kv,t$-space, then this relation must hold for all $\vec{k}$, $t$. 
By ordering the indices such that $n>m$ implies $\phi_n\geq \phi_m$, this prescription defines a unique labeling of the phase bands.

The ordering condition above is constructed such that 
if two phase bands become degenerate at a particular value of $\vec{k},t$, the bands do not ``cross'' (e.g., the index $m$ stays with the lower branch everywhere, while the index $n$ stays with the upper branch). 
This arrangement is in particular maintained when a band in the fundamental zone meets a band originating from a different zone, 
see for example band \#3 in Fig.~\ref{fig: Deformation 1}d. 
Such degeneracies between phase bands associated with different branches play an essential role in defining the topological characteristics that distinguish driven and non-driven systems.



\begin{figure}\begin{center}
\includegraphics[scale=0.58]{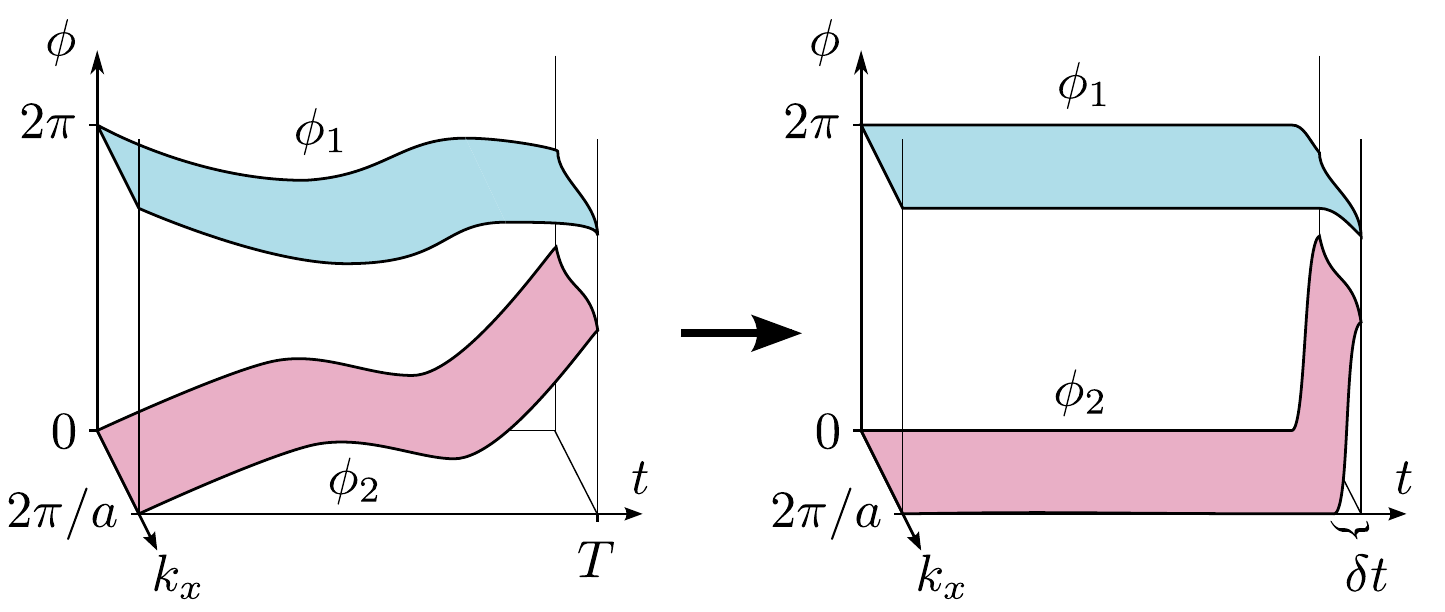}
\caption{Graphical depiction of the deformation described in Sec.~\ref{sec:PhaseBands}, where the time-evolution operator of a periodically driven system is deformed into the time-evolution operator of a non-driven system. This deformation is always possible if the time-evolution operator has no degeneracies.
After flattening, the linear ramp region is expanded to the entire interval $0<t<T$, and the bands are straightened into a form as in Fig.~\ref{fig: Deformation 1}a.}
\label{fig:TwoStep}
\vspace{-0.2in}
\end{center}\end{figure}
We now use the phase band picture to demonstrate when it is possible to continuously deform the evolution of a Floquet-Bloch system into that of a non-driven system, while keeping $U(\vec{k}, T)$ fixed. 
%
Naively, the continuity of $U(\kv,t)$ and of the phase bands might lead one to expect that the projectors $P_n(\kv,t)$ are continuous as well. 
If this were true, any continuous deformation of the phase bands $\phi_n(\vec{k},t)$ would preserve the continuity of the evolution operator. 
It would then always be possible to deform the evolution into that of a non-driven system using a two-step ``band-flattening'' procedure (see Fig.~\ref{fig:TwoStep}).
First, for every $n = 1\ldots N$, deform $\phi_n(\kv,t)$ to zero for all $0 \le t \le T - \delta t$ until a small time-interval $\delta t$ before $T$, after which 
it grows linearly to its final value. 
If the interval $\delta t$ is small enough, we can assume that the projectors are constant there, $P_n(\kv,t)=P_n(\kv,T)$. 
In the second step, let $\delta t \rightarrow T$, while keeping the projectors constant 
throughout the linear ramp of the phase. 
The deformed evolution is now identical to that of a non-driven system with the Hamiltonian 
\be 
h(\kv)=\frac 1 T \sum_n \phi_n(\kv,T) P_n(\kv,T).
\label{eq: effective Hamiltonian}
\ee

The picture above seems to imply that all periodically-driven systems are topologically equivalent to non-driven systems (i.e., they can be related by smooth deformations that keep the Floquet operator fixed). 
However, the existence of phenomena such as anomalous edge modes\cite{Winding_Number, KBRD} 
shows that this {\it cannot} be the case. 

Where could the argument break down?
In the first step, we assumed that the phase bands could be continuously deformed to zero throughout the entire driving period, up to a short interval $\delta t$ in which the projectors were assumed to be constant.
However, in principle one may imagine that the evolution operator could 
host degeneracies around which the projectors 
are {\it discontinuous} (the degeneracy of the eigenvalues ensures that $U$ stays continuous). 
In the presence of such a discontinuity, the degeneracy could not be lifted without breaking the continuity of $U$. 
In this way a phase band in the fundamental zone may become ``glued'' to  another band from a neighboring branch of phases (see Fig.~\ref{fig: Deformation 1}c). 

In the absence of the discontinuities described above,  the evolution of any driven system can be smoothly deformed to that of a non-driven system, as in Fig.~\ref{fig:TwoStep}, and anomalous edge states would be impossible.   
Thus we are led to the unavoidable conclusion that the evolution operators of periodically driven systems must support topologically-protected degeneracies. 
In the next subsection we show explicitly 
that such degeneracies can exist 
in two-dimensional (2D) systems.
In Sec.~\ref{cha:SYM} we generalize to other dimensions and symmetry classes.
Below we refer to these topologically-protected degeneracies as ``topological singularities.''

\subsection{Topological singularities in two-dimensional systems}
\label{sec: Topological singularities in 2D systems}
\label{sec:TOP_singularities}

In this subsection we explicitly demonstrate the existence and nature of 
topological singularities in the evolution operators of two-dimensional systems.
We furthermore show that in a region in $\kv,t$-space where $U(\kv,t)$ is degenerate, the degeneracy can either be lifted everywhere or reduced to a cluster of isolated 
singularities.

Let $U(\kv,t)$ be the bulk time evolution operator of  a two-dimensional system with no symmetries other than the discrete translational symmetry of the lattice. 
Consider now a point $\vec{s}_0=(\kv_0,t_0)$ in $\kv,t$-space where 
two adjacent phase bands, 
 $m$ and $m'$,
are degenerate (mod $2\pi$).
At $\vec{s}_0$, the degenerate subspace is spanned by the Bloch states 
$|\psi_m\rangle$ and $|\psi_{m'}\rangle$.
Due to the continuity of $U(\vec{k},t)$ and the existence of gaps to other phase bands, 
 we can assume that the subspace spanned by the 
two intersecting bands is constant within some finite sized neighborhood around $\vec{s}_0$ in $\kv,t$-space. 
The remaining non-degenerate bands $\{|\chi_n\rangle\}$ and their associated phases $\{\phi_n\}$ can also be assumed to be constant within this neighbourhood. 
Close to $\vec{s}_0$, the time evolution operator thus takes the form
\be 
U(\vec{s}) =\!\!\! \sum_{n\neq m,m'} 
|\chi_n\rangle \langle \chi_n| e^{-i\phi_n} 
+  \!\!\! \sum_{a,b=m,m'}
|\psi_a\rangle \langle \psi_b|M_{ab}(\vec{s}),
\label{eq: U form}\ee
where $M$ is a $2\times 2$ unitary matrix and we parametrize the three-dimensional $(\vec{k},t)$ space by a single variable $\vec{s}$. 

The unitarity of $M$ 
means that we can write it as
\be
M(\vec{s})= \exp\left[-i\phi_d(\vec{s})-i f_j(\vec{s}) \sigma_j \right],
\label{eq: M form1}
\ee
where summation over repeated indices is used. Here $\phi_d(\vec{s})$ is a real-valued function, whose value at $\vec{s}_0$ gives the common eigenvalue of the two degenerate bands, 
$\{\sigma_j\}$ are the Pauli matrices, and $\{f_j(\vec{s})\}$ are real continuous functions that satisfy $f_i(\vec{s}_0)=0$.  

We assume that $U$, and thereby $f$, is differentiable in a neighborhood around $\vec{s}_0$, and expand $f_j$ to linear order in $(\vec{s}-\vec{s}_0)$ around $\vec{s}_0$. 
Noting that $f_j(\vec{s}_0)=0$, we write 
\be 
M(\vec{s}) \approx \exp\left[-i\phi_d(\vec{s})-i(\vec{s}-\vec{s}_0)_jS_{jk} \sigma_k\right], 
\label{eq: M form}
\ee
 where $S_{jk}=\partial_j f_k(\vec{s}_0)$ is a real $3\times 3$ matrix.  
The case where the linear term in $(\vec{s}-\vec{s}_0)$ also vanishes 
will be covered shortly.

We first consider the case where the matrix $S$ has rank three, such that the coefficients of all three Pauli matrices vary independently as $\vec{s}$ explores the neighborhood around $\vec{s}_0$. 
In this case, the degeneracy is  topologically protected, similar to the case of a Weyl node~\cite{Weyl}: an infinitesimal change of the time evolution operator can never lift the degeneracy, but rather can only infinitesimally shift the location where it appears. 
A single such degeneracy can thus not be lifted with a continuous deformation of the system, and is therefore topologically protected. 
We thus define 
a topological singularity of a two-dimensional system to be an isolated degeneracy of the time evolution operator where the matrix $S$ describing the linearization of $U$ in its neighborhood [Eq.~(\ref{eq: M form})] is invertible. 

In addition to the isolated singularities described above, we may also find cases where $S$ is not invertible. 
This occurs when two phase bands are degenerate along a line, surface, or three-dimensional region in $\kv,t$-space, such that $\vec{s}_0$ is one point on this manifold. 
In such cases, the rank of $S$ is equal to 
$3-D$, where $D$ is the dimension of the degenerate manifold. 
These extended degeneracies are not topologically protected: 
the degeneracy can generically be lifted in a neighborhood of $\vec{s}_0$ with a local perturbation, letting $f_i(\vec{s})\rightarrow f_i(\vec{s})+\delta g(\vec s) v_i$ in Eq.~(\ref{eq: M form1}). 
Here $\delta$ controls the strength of the perturbation, $\vec{v}$ is a 3-dimensional vector satisfying $v_iS_{ij}=0$, and $g(\vec s)$ is a real continuous function that vanishes outside  a neighbourhood around $\vec s_0$, within which $|\psi_{m,m'}\rangle$ can be taken to be constant.

Importantly, the local perturbations described above only lift the degeneracy patch-wise, in one small region at a time. 
If one tries to lift the degeneracy over the entire manifold, two cases are possible: either the degeneracy can be lifted everywhere, or there will be a discrete set of points where topological singularities remain. 
Hence we conclude that, if the time evolution operator is degenerate 
throughout a finite-dimensional manifold, 
it is always possible to apply an infinitesimal 
perturbation that either completely lifts the degeneracy, or reduces it to a cluster of topological singularities.

With the existence of topological singularities established, we now further characterize their properties. 
Each singularity can be assigned a charge (or vorticity) $q$:  
\be 
q = \sgn\, [\det S], 
\label{eq: singularity charge definition}
\ee
where $S$ is the linearization of $f_k(\vec{s})$ around $\vec{s}_0$, see Eqs.~(\ref{eq: M form1}) and (\ref{eq: M form}).
%
%
%

In two-dimensional systems, the charges of topological singularities have direct connections with the Chern numbers of the phase bands. 
Consider 
the ``instantaneous'' Chern number of phase band $n$, $C_n(t) = \frac{1}{4\pi}\int d^2k\, {\rm Tr}\{P_n[ \partial_{k_x}P_n, \partial_{k_y}P_n]\}$. 
As long as no singularities are encountered, the Chern number $C_n(t)$ cannot change due to the continuity of $P_n(t)$. 
 However, when  two phase bands meet at a singularity  with vorticity $q$, the Chern number for the ``upper'' phase band 
changes by $q$ as the singularity is traversed in time, while 
the Chern number of the other band changes by $-q$.
Here, the ``upper'' band is band $m+1$ if the singularity connects bands $m$ and $m+1$, and band $1$, if the singularity connects band $1$ and $N$ through the phase Brillouin zone edge.
As a consequence of the argument above, any driving protocol that yields Floquet bands with different Chern numbers from those of the initial Hamiltonian $H(0)$ {\it must} induce one or more topological singularities in $U(\vec{k},t)$.

In this section we showed that the evolution operator of a periodically-driven system may host topologically-protected degeneracies, or ``topological singularities.'' 
As we concluded in the  beginning of the section, topological singularities can obstruct the smooth deformation of the evolution of the driven system into that of a non-driven system.
Specifically, in the case where the ``bottom'' and ``top'' phase bands, $1$ and $N$, are connected by a singularity through the phase Brillouin zone-edge, their respective phase values at the singularity must differ by $2\pi$. 
In this situation it is impossible to simultaneously flatten both of the bands to zero
(compare Figs.~\ref{fig: Deformation 1}c, d with Fig.~\ref{fig:TwoStep}).
In contrast, for singularities that {\it do not} pass through the phase zone edge (i.e., those connecting bands $m$ and $m+1$, with $m < N$), the two corresponding phase eigenvalues coincide at the singularity.  
In this case nothing prevents deforming the two phase bands simultaneously to zero, thereby removing the singularity.
Hence 
singularities  of the first type, i.e., ``zone-edge singularities,'' are special: it is precisely these singularities
that cannot be eliminated by smooth phase-band deformations, thus distinguishing driven from non-driven evolution.
In Sec.~\ref{sec:2dCase} below we formulate the topological classification for two-dimensional periodically driven systems in terms of the phase bands and zone-edge topological singularities, and derive the corresponding bulk-edge correspondence.

\subsection{Natural quasienergy zone}
\label{sec:NaturalQuasiEnergyZone}
Before moving on to classification, we briefly introduce some further labeling notation that will be useful for referring to specific Floquet bands and gaps in the discussion below.
In particular, we apply the phase band labeling prescribed in the beginning of this section 
to the Floquet bands themselves. 
We label Floquet band $n$ such that $\varepsilon_n(\kv)=\phi_n(\kv,T)/T$ (see Fig.~\ref{fig: Deformation 1}d).
This uniquely determines ``natural'' quasienergy band indices 
for a system, and the bands $\{\varepsilon_n\}$ for $n=1\ldots N$ define a natural quasienergy (or Floquet) Brillouin zone.

The Floquet band indices 
are topologically invariant: 
within the predefined labeling scheme, the index of a Floquet band cannot change under any smooth deformation that keeps the gaps above and below 
the band open.
Notably, the natural indices of the Floquet bands can be found directly with the help of the time-averaged Hamiltonian, without having to track the full time development of the phase bands. 
In particular, note that
\be 
\label{eq:avgH} \sum_n\varepsilon_n(\kv)=\frac{1}{T}\int_0^T\! dt\, \Tr\left[H(\kv,t)\right].
\ee
Recall that all  $\varepsilon_n(\kv)$ lie within an interval of width $2\pi/T$, and that the bands are ordered by increasing $n$.
From this information and Eq.~(\ref{eq:avgH}) above, the indexing of Floquet bands is uniquely determined. 

Below we will also apply the labeling scheme to the quasienergy {\it gaps}. 
For the following discussion, we refer to  the quasienergy gap {\it above} band $m$ as gap $m$.
Due to the periodicity of quasienergy, a driven system with $N$ bands has an additional gap (as compared to a non-driven system), which separates band $N$ from band 1, across the quasienergy zone edge.
We thus refer to gap $N$ as the ``zone-edge gap'' of the Floquet spectrum.
In the non-driven limit 
$T\rightarrow 0$, 
the zone-edge gap becomes infinitely wide, while the other gaps remain finite.

\section{Topological classification of Floquet-Bloch systems in two dimensions} 
\label{sec:2dCase}
\begin{figure}\begin{center}
\includegraphics[scale=0.58]{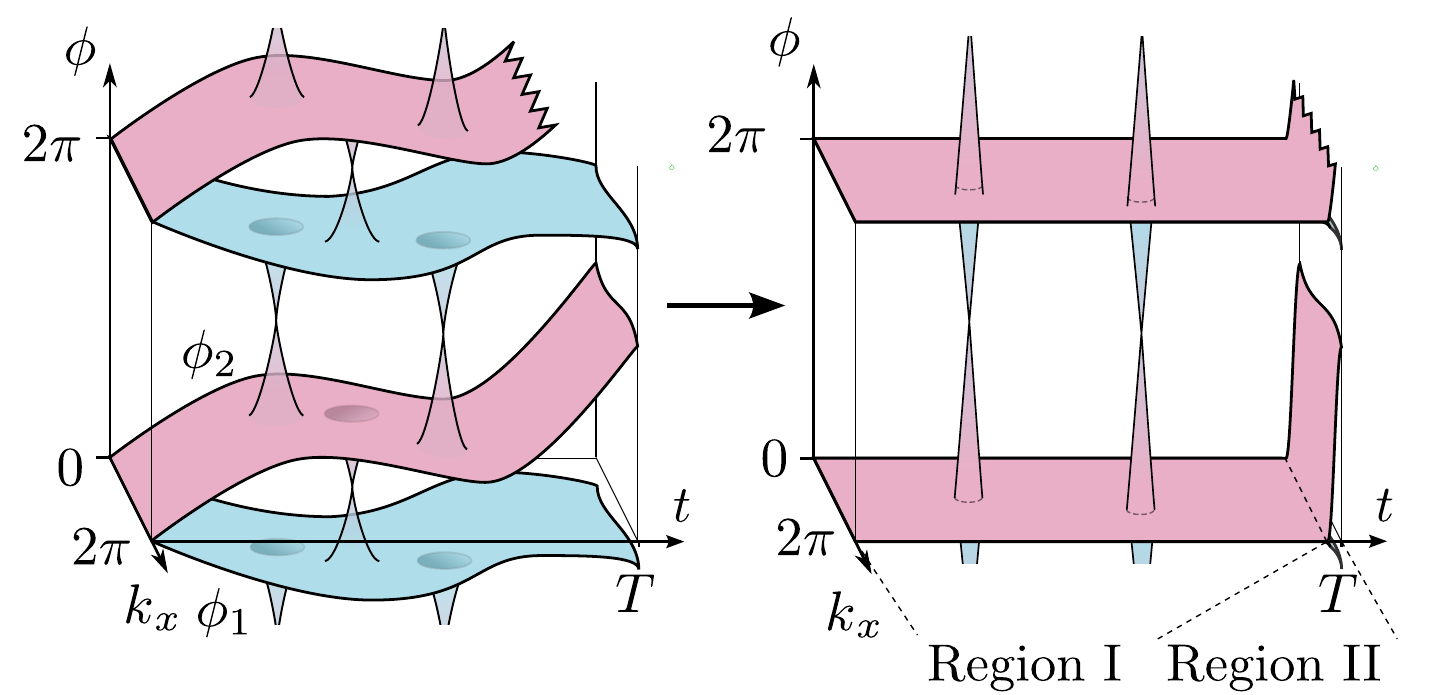}
\caption{
Continuous deformation of a generic time evolution operator which only preserves information about the topological singularities spanning the 
zone-edge gap (region I), and the Floquet bands (region II).
These features cannot be removed, since the Chern indices of the Floquet bands and the total charge of the zone-edge singularities can only be changed by closing at least one quasienergy gap. 
Any topological invariant should be expressible in terms of only these non-removable features.
 Note that the singularity in the gap around phase $\phi = 0$ is eliminated under the deformation.  
}
\label{fig:NSC:Singularity_graph}
\end{center}\end{figure}
Having introduced the concept of phase bands and demonstrated the existence of topological singularities, we now consider the implications of these results for the topological properties of a two-dimensional Floquet-Bloch system with no  symmetries.
By ``topological''  we mean those properties 
 that are invariant under any continuous deformation of the bulk time-evolution operator that  preserves its continuity in crystal momentum and time, and keeps the bulk gaps open in the quasienergy spectrum of the Floquet operator $U(T)$. 
Any such quantity is a topological invariant of the system.
Importantly, this definition means that topological invariants must be independent of the 
choice of time origin\cite{footnote:time_origin}. 

Analogous to a Chern insulator, a two-dimensional Floquet-Bloch system defined in a geometry with edges may host protected chiral edge modes within its bulk quasienergy gaps.
The chiral edge modes are topologically protected, meaning that the net number $\nedge (m)$ of chiral edge modes in  bulk quasienergy gap $m$ is invariant under continuous deformations of the bulk evolution operator $U(\kv,t)$, or equivalently of the Hamiltonian $H(t)$, that keep quasienergy gap $m$ open. 
Thus we recognize $\nedge(m)$ as a topological invariant of the system. 

In this section we demonstrate that considerations about the bulk phase bands
allow us to identify all independent topological invariants of a two-dimensional Floquet-Bloch system.
Subsequently, we  use these invariants to construct the bulk-edge correspondence, providing a direct link between the edge mode spectrum and the bulk phase band properties of two-dimensional Floquet-Bloch systems. 

\subsection{{Topological invariants of 2D  systems}} 
\label{sec:NSC:Complete_classification}

In this subsection we identify the properties of a two-dimensional Floquet-Bloch system  that are invariant under smooth deformations of the phase bands.
In Sec.~\ref{sec:TOP_singularities} we found that the phase bands of a 2D system can safely be 
deformed anywhere in $\kv,t$-space except at isolated topological singularities.  
Therefore we may expect that the singularities play an important role in the topological classification of periodically driven systems.

To elucidate the importance of topological singularities we consider the following deformation of a time-evolution operator $U(\vec{k},t)$, shown schematically in Fig.~\ref{fig:NSC:Singularity_graph}. 
Without changing the eigenstate projectors $P_n(\kv,t)$, 
deform the phase bands $\phi_n(\vec{k},t)$ to zero everywhere except for within small isolated regions that surround each zone-edge singularity 
and within a small time-interval $\delta t$ before $T$, where 
the phase bands wind linearly to their final values. 
These final values, which define the Floquet bands, are  kept fixed under the deformation.
{We refer to the region $t<T-\delta t$ in $\kv,t$ space as region I, and to the final region $t>T-\delta t$ as region II (see Fig.~\ref{fig:NSC:Singularity_graph}).
Any time-evolution operator can be deformed in this way such that continuity is preserved and no quasienergy gap is closed. 
Without changing any topological invariant, the deformation effectively discards all information about the time-evolution operator except for  the phase bands at time $t=T$ (i.e., the Floquet bands themselves), and the zone-edge singularities.

Consider now the remaining features of the phase band structure that could not be smoothly deformed away. 
We found in Sec.~\ref{sec: Topological singularities in 2D systems} that it was possible to change the location $\kv,t$ of each singularity through a continuous deformation.
Through such a deformation, it is furthermore possible to create and annihilate pairs of zone-edge singularities with opposite charges. 
Hence the only invariant quantity we can associate with region I is the sum of the charges $\{q_i^{\rm (ZES)}\}$ of all zone-edge singularities, $\sum_i q_i^{({\rm ZES})}$. 
For region II, we note that at $t=T$ any two projectors $P(\kv,T)$ and $P'(\kv,T)$ can be continuously deformed into each other if and only if their Chern numbers are the same\cite{AvronPRL1983}. 
Hence the only independent invariants we can associate with region II are the Chern numbers of the individual phase bands at $t=T$.
 
The arguments above show that a two-dimensional Floquet-Bloch system with $N$  bands 
has exactly \textit{$N$} independent topological invariants characterizing it. These invariants are the integers
\be 
\left( C_1 , \ldots  C_{N-1},
\sum_i q_i^{\rm (ZES)}\right),
\label{eq:NSC:NscInvariants}
\ee
where ${C}_n$ is the Chern number of Floquet band $n$ (see Sec.~\ref{sec:NaturalQuasiEnergyZone} for definition of the quasienergy band indices). 
The index $i$ in the sum runs over all topological singularities in the zone-edge gap.   
The Chern number of the last band $ C_N$ is not included since $\sum_n  C_n=0$. 
We see that while an $N$-band  non-driven system is characterized by $N-1$ independent integer-valued ($\mathbb{Z}$) invariants 
(the Chern numbers of each of the $N-1$ lowest bands),
Floquet-Bloch systems are characterized by $N$  integer ($\mathbb{Z}$) topological invariants. The additional invariant is the net charge of the  topological singularities in the zone-edge gap. 

%

\subsubsection{Bulk-edge correspondence for two-dimensional Floquet-Bloch systems}
\label{sec:Bulk-edge_Correspondence_2D}

%
%
%
%

We now seek to derive a bulk-edge correspondence that gives the net number of chiral edge states that will appear within a given gap $m$ of the bulk Floquet spectrum when the system is defined in a geometry with an edge.
To this end we seek the} combinations of the $N$ numbers in Eq.~(\ref{eq:NSC:NscInvariants}) that remain invariant when all gaps except for gap $m$ are allowed to close (see Sec.~\ref{sec:NaturalQuasiEnergyZone} for the labeling convention for the quasienergy gaps).
We know at least one such quantity should exist since the number $\nedge(m)$ of edge modes in gap $m$ has this property.
%

In order to find the invariant combinations, we note that the Chern numbers of the individual bands $1$ to $m$ can be changed by closing the quasienergy gaps between them. 
Only their sum $S_m = \sum_{n=1}^m C_n$ remains constant under such operations. 
Furthermore, if $m\neq N$, all zone-edge singularities can be removed through the plane $t=T$ by closing the zone-edge quasienergy gap (i.e., the gap between band $N$ and band $1$). 
Importantly, however, the Chern number of band $1$,  and thereby $S_m$,  changes by $q$ each time a singularity of charge $q$ is removed in this way (see the discussion at the end of Sec.~\ref{sec:TOP_singularities} on the relationship between Chern numbers and singularities). 
Hence there only exists \textit{one} combination of the invariants in Eq.~\eqref{eq:NSC:NscInvariants} which remains invariant when only gap $m$ is required to stay open:
\be  
w_m[U]=\sum_{n=1}^m  C_n - \sum_i q_i^{\rm (ZES)}.
\label{eq:NuMInvariant}
\ee

Any two evolutions characterized by the same value of the invariant $w_m$ can be smoothly deformed into one another without closing quasienergy gap $m$.
Crucially, this tells us that the number of chiral edge modes $\nedge(m)$ in gap $m$ should be some function of $w_m$, and possibly $m$ itself.
Standard spectral flow arguments require $\nedge(m)-\nedge(m-1)=C_m$. The only way this can be realized is if $\nedge(m)=w_m + K$ for some universal constant $K$. 
Considering the trivial special case $H(t)=0$, where both $w_N$ and $\nedge(N)$ are zero, we find that $K$ must be zero.
We thus arrive at the following new result for the net number of chiral edge modes in a two-dimensional system:
\be 
\nedge (m)=
\sum_{n=1}^m  C_n - \sum_i q_i^{\rm (ZES)}.
\label{eq: Winding number sum}
\ee

The simple expression above provides a direct way of evaluating the edge mode count given by the winding number formula found in Ref.~\onlinecite{Winding_Number}. 
The first term is the result one obtains simply when analyzing a non-driven system with the phase band framework, 
taking $T$ to be so small that the phase bands do not cross.
The second term has no equivalent in non-driven systems, and accounts for the anomalous edge modes that were discussed in Refs.~\onlinecite{KBRD, Winding_Number}. 
Additionally, Eq.~(\ref{eq: Winding number sum}) shows that the number of edge modes in the zone-edge  gap is given by the net charge of all zone-edge topological singularities.

In Appendix~\ref{app:Time-domain_Expressions_NSC} we provide an explicit derivation showing that Eq.~\eqref{eq: Winding number sum} is equivalent 
to the winding number formula of Ref.~\onlinecite{Winding_Number}.
Below we refer to $w_m[U]$ as the winding number of $U$ in gap $m$.

%
%

\subsection{Topological singularities in a specific 2-band model}
\label{sec: Specific Model}

\begin{figure*}\begin{center}
\includegraphics[scale=0.58]{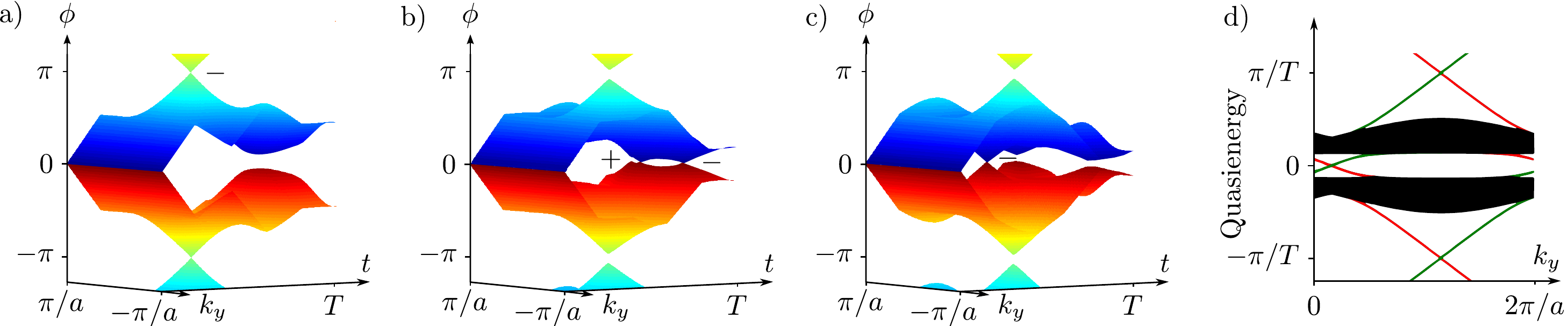}
\caption{Explicit demonstration of topological singularities and anomalous edge states. 
a-c) Phase band structures for the model in Eq.~(\ref{eq: Model }) for fixed values of $k_x$, with singularity charges (all $\pm 1$) indicated. 
The $k_x$ values are a) $0.245/a$, b) $1.533/a$, c) $2.084/a$. 
d) Quasienergy band structure of the model in a strip geometry. 
Both bands have Chern number zero, and we find one chiral mode on each edge, in each quasienergy gap.  
Edge modes on opposite edges are indicated by different colors.} 
\label{fig: Band structure plot}
\vspace{-0.2 in}
\end{center}\end{figure*}

To make our discussion more concrete, in this subsection we demonstrate the results above on a variation of the explicit model considered in Ref.~\onlinecite{Winding_Number}.
Consider a tight-binding model on a 2D bipartite square lattice, described by the time-dependent Bloch Hamiltonian 
\be 
H(\kv,t)=\sum_{n=1}^4 J_n(t)\left(\sigma^+ e^{i \mathbf b _n \cdot \kv}+ \sigma^- e^{-i \mathbf b_n \cdot \kv} \right)+ V \sigma_z,
\label{eq: Model }\ee
where $\sigma_z$ and $\sigma^\pm=(\sigma_x\pm i \sigma _y) / 2$ are the Pauli matrices acting in the sublattice space, and the vectors $\{\mathbf b_n\}$ are given by $\mathbf b_1 = -\mathbf b_3 =(a,0)$, and $\mathbf b_2= - \mathbf b_4 = (0,a)$, with $a$ being the lattice constant.
In real-space, Hamiltonian (\ref{eq: Model }) 
consists of hopping terms between nearest neighbour sites on the bipartite lattice.
The Hamiltonian is $T$-periodic in time. 
Each driving cycle consists of five time intervals of length $T/5$, with $J_n(t)=\lambda_n$ during the $n$-th interval, while all the other hopping amplitudes are set to zero.
In the fifth interval, all hopping amplitudes are zero while the sublattice potential $V$ remains on.

In Ref.~\onlinecite{Winding_Number}, anomalous edge modes were observed in  the case where $\lambda_n=J$, for certain ranges of the parameters $J$ and $V$.
According to the discussion in the subsections above, this implies that topological singularities are present. 
Indeed, when in a nontrivial phase, the two phase bands touch through the zone-edge along the line $k_x=k_y$, at a specific time that depends on 
parameter values. 
To demonstrate that this degenerate region contains topological singularities, we add a small time-dependent perturbation to break the extended degeneracy into isolated singular points 
(see section \ref{sec: Topological singularities in 2D systems}). 
We implement the perturbation by reducing the hopping in the $y$-direction slightly compared to the $x$-direction, such that  $\lambda_1=\lambda_3 = J$ and  $\lambda_2=\lambda_4=(1-\alpha) J$, where $\alpha$ is a small parameter. 
We then  numerically calculate the time-evolution operator at a representative set of points in $(\kv,t)$-space for the parameter choice $J=-2.5 \pi / T$, $V=0.8 \pi /T$, and $\alpha = 0.2$.   
From diagonalization of the time-evolution operator we obtain the phase band structure of the model, and find four topological singularities (see Fig.~\ref{fig: Band structure plot}).
One singularity has charge $-1$ and 
connects the two bands through
the zone edge, while the other three have charges $1,-1$, and $-1$, but do not cross 
the zone-edge.
The charges are found numerically. 

In Figs.~\ref{fig: Band structure plot}a-c, the phase band structure is plotted for three values of fixed $k_x$. 
The $k_x$ values are chosen where the four topological singularities appear (two of the singularities appear at the same $k_x$). 
The Chern numbers of the Floquet bands are zero. 


Next we confine the model to a strip geometry with edges parallel to the $y$-direction, by truncating the 
real-space 
Hamiltonian of the model 
in the $x$-direction.
We numerically calculate the Floquet operator of this truncated tight-binding 
Hamiltonian and obtain the quasienergy band structure shown in Fig.~\ref{fig: Band structure plot}d. 
On each edge we find the net number of chiral edge modes to be 1, in both bulk quasienergy gaps.
This behavior is fully consistent with result (\ref{eq: Winding number sum}) above.

\section{Topological classification of Floquet-Bloch systems with symmetries}
\label{cha:SYM}
In the previous section we showed that the richer topological structure of two-dimensional periodically driven systems was due to the possibility of non-removable singularities in the phase bands of such systems.
%
Building on this result, we now seek to describe how additional restrictions on the evolution (e.g., as imposed by discrete symmetries) can protect new types of phase band singularities in one, two, or three dimensions.
These new singularities provide the basis for a symmetry-based topological classification of Floquet-Bloch systems.

Inspired by the rich structure of the periodic table of topological insulators in non-driven systems\cite{Periodic_Table_2, Periodic_Table_1}, we focus on driven system analogues of the ten Altland-Zirnbauer (AZ) symmetry classes.
In the first subsection below we describe two types of symmetry conditions on the evolution operator (``instantaneous'' or ``time non-local'') which provide useful ways of generalizing the AZ symmetries to driven systems. 

Note that the instantaneous and time non-local conditions are chosen as illustrative examples to demonstrate the power and adaptability of the phase band framework. 
These conditions are not necessarily the only ways of generalizing the AZ symmetries.
Note also that these are not the only types of conditions that can protect singularities -- it will be an interesting direction for future work to seek wholly new types of symmetries on $U(t)$ which may protect additional types of topological singularities. 

Following the discussion of smooth phase band deformations from Sec.~\ref{sec:2dCase}, we 
find the exhaustive classification for one-dimensional systems with particle-hole symmetry and identify the related bulk-edge correspondences. 
We then go on to find the bulk-edge correspondences for two- and three-dimensional systems with time-reversal symmetry. 
The bulk-edge correspondences that we obtain in this section for one- and two-dimensional systems coincide with those found in Refs.~\onlinecite{1D_Majorana, 2D_TR}, respectively.
To our knowledge, analogous results for the three dimensional case have not been derived before. 

Within the symmetry framework that we consider, we find that the edge mode spectrum in a single bulk Floquet gap has the same classification for driven and non-driven systems. 
However, the \textit{global} edge mode spectrum of a periodically driven system generally has a {\it richer} classification than its non-driven counterpart, with a correspondingly richer mathematical relationship between bulk and edge properties.
For each symmetry class, we thus find a larger number of distinct topological phases than in the corresponding non-driven cases.
Interestingly, we find for example that a periodically driven \textit{two-band} system with time-reversal symmetry can host topologically protected edge modes, while a minimum of four bands is necessary in the non-driven case.
This richer variety of topological phases in driven systems originates from the periodicity of quasienergy, i.e., the presence of the zone-edge gap in the Floquet spectrum, and the existence of symmetry-protected topological singularities which may reside 
in the corresponding phase band gap.

\subsection{Symmetries in periodically driven systems}
\label{sec:Symmetries intro}
We now identify two symmetry conditions on the evolution operator which can protect new types of singularities in the ``time-bulk'' of the phase band structure (i.e., singularities occurring for intermediate times $0 \le t < T$): 
``instantaneous'' symmetries of the form
\be
\label{eq:instantaneous}
U(t)=SU(t)S^{-1},
\ee 
and ``time non-local'' symmetries of the form
\be 
\label{eq:timenonlocal}U(t)=SU(t^*-t)U^\dagger(t^*) S^{-1}.
\ee
Here $S$ may be 
a unitary or an anti-unitary operator, and $t_*$ denotes a special point in the driving cycle. 
In the subsequent discussion, we always pick the time origin such that $t_*=T$.
Note that the instantaneous symmetries 
relate the evolution operator to itself at a given time, while the time-non-local symmetries 
relate the evolution operator to itself at {\it different} times.


The conditions in Eqs.~(\ref{eq:instantaneous}) and (\ref{eq:timenonlocal}) can  be used to ensure that the Floquet operator $U(T)$ and/or the corresponding ``effective Hamiltonian'' $H_{\rm eff}$, defined via $U(T) = e^{-i H_{\rm eff}T}$, falls into any one of the ten Altland-Zirnbauer symmetry classes.
For example, particle-hole symmetry is guaranteed via an instantaneous condition of the form (\ref{eq:instantaneous}), with $S$ anti-unitary (see below).
On the other hand, time-reversal and chiral symmetries are imposed 
via time-non-local conditions of the form (\ref{eq:timenonlocal}), with $S$ anti-unitary and unitary, respectively.
Furthermore, just as in the non-driven case, the symmetry conditions where $S$ is anti-unitary divide into two subclasses,  depending on whether $S$ squares to $1$ or $-1$.


\subsection{Particle-hole symmetry with $S^2=1$}
\label{sec:PHS}
In this subsection we use the phase band framework to develop a topological classification for periodically driven systems with particle-hole symmetry (PHS). 
Here we impose PHS via an instantaneous symmetry condition as in Eq.~\eqref{eq:instantaneous}, where the operator $S$ is anti-unitary and  squares to $1$ (analogous to symmetry class D in the AZ convention\cite{AZ_Convention}).
Such a condition is naturally satisfied, for example, in the Bogoliubov-de Gennes Hamiltonian of a driven spinless superconductor.
The one-dimensional (1D) case was considered previously in Ref.~\onlinecite{1D_Majorana}.
Here we use the phase band framework to identify topological invariants and to derive the bulk-edge correspondence, 
obtaining results consistent with the findings of Ref.~\onlinecite{1D_Majorana}.

The condition above implies that ``class D type PHS'' is present if and only if there exists a basis where the instantaneous Bloch Hamiltonian of the driven system, $h(\kv,t)$,  satisfies  $h(\kv,t)=-h(-\kv,t)^*$.
Consequently, in this basis, the evolution operator  at each time $t$ satisfies $U(\kv,t)=U^*(-\kv,t) $.
This furthermore implies that the time evolution operator of a particle-hole symmetric system with $2N$ bands can be written in the form
\be 
U(\kv,t)=\sum_{n=1}^N\left[P_n(\kv,t)e^{-i\phi_n(\kv,t)}+P_{\bar n}(\kv,t)e^{-i\phi_{\bar n}(\kv,t)}\right],
\label{eq:Particle-hole symmetric time-evolution operator}
\ee
where the phase bands $\{\phi_n(\kv,t),\phi_ {\bar n}(\kv,t)\}$ are  continuous and non-crossing (as defined in Sec.~\ref{sec:PhaseBands}), and, in the basis specified above,
\begin{align}
P_{\bar n}(\kv,t)=P^*_n(-\kv,t), \quad
\phi_{\bar n}(\kv,t)=-\phi_n(-\kv,t).
\label{eq:PHS:conjugate band relation}
\end{align}  
The ambiguity of the labeling of bands is removed by requiring $0\leq \phi_n\leq \pi$, with the conjugate phases satisfying $-\pi \le \phi_{\bar{n}} \le 0$.
In previous sections we labeled the bands according to increasing $n$, starting from the lowest band. 
Here, making use of the symmetry of the spectrum, we start the labeling from the band with the smallest positive phase.

Similar to the approach in Secs.~\ref{sec:PhaseBands}~and~\ref{sec:2dCase}, we now seek to identify topological invariants by considering quantities that do not change under smooth deformations of the evolution {\it which preserve the particle-hole symmetry}, expressed via Eqs.~(\ref{eq:Particle-hole symmetric time-evolution operator}) and (\ref{eq:PHS:conjugate band relation}).
Preservation of the symmetry can be ensured by continuously deforming {\it half} of the phase bands and 
projectors, $\{\phi_n(\kv,t),P_n(\kv,t)\}$, with their conjugate partners following in accordance with condition (\ref{eq:PHS:conjugate band relation}).
However, as we found for the two-dimensional case considered in the previous section, free deformation of the phase bands may be obstructed at certain isolated points in $\kv,t$-space where protected singularities are encountered. 

In order to see how a singularity may be protected by particle-hole symmetry (in any dimension), 
consider an inversion invariant point $\kinv$ 
in the Brillouin zone, 
where Eq.~(\ref{eq:PHS:conjugate band relation}) directly relates the phases and projectors of conjugate partner bands. 
Suppose that at some time $t_0$ two conjugate phase bands $m$ and $\bar m$ 
become degenerate  at $\kinv$; 
with the labeling we prescribed earlier, $m$ may be either $1$ or $N$.
Because these bands form a conjugate pair, the degeneracy may only occur at phase $0$ or $\pi$, see Eq.~(\ref{eq:PHS:conjugate band relation}).
For times $t$ close to $t_0$, the evolution operator at $\kinv$
 can be written 
\begin{eqnarray}
U(\kinv,t)&=&\sum_{n\neq m}\left( |\chi_n\rangle\langle \chi_n| e^{-i\phi_n(t)}+|\chi_n^*\rangle\langle\chi_n^*| e^{i\phi_n(t)}\right)\notag\\
&+& \sum_{a,b=1,2}M_{ab}(t)\, |\psi_a\rangle \langle \psi_b| ,
\label{eq:PHS singularity}
\end{eqnarray}
where $\{|\chi_n\rangle,|\chi_n^*\rangle\}$ are the eigenstates of $U(\kinv,t_0)$ that do not become degenerate (assuming the system has more than two bands), and $\{|\psi_{1,2}\rangle\}$ are two states that together span the subspace of the degenerate eigenstates $m$ and $\bar m$ at $t = t_0$.
The $2\times 2$ matrix $M(t)$ is  unitary and depends continuously on $t$. 
Furthermore it satisfies $M_{ab}(t_0)= \pm \delta_{ab}$, where the sign depends on whether the bands meet at phase $\pi$ (for $-$) or at phase $0$ (giving $+$).

The symmetry (\ref{eq:instantaneous}) 
implies that $U(\kinv,t)$ must be real for all $t$. 
Furthermore,  we can take 
$|\psi_1\rangle,|\psi_2\rangle$ to be real, since the two degenerate eigenstates $\{|\chi_m\rangle,|\chi_{\bar{m}}\rangle\}$ are complex conjugates of each other.
The first sum in  Eq.~(\ref{eq:PHS singularity}) is also real, thus implying that $M(t)$ must itself be real. 
As a result, we can write $M(t)$ as
\be 
M(t)=\pm e^{-i \lambda \sigma_{y} (t-t_0)},
\label{eq:PHS singularity 2}
\ee
where the parameter $\lambda$ is real. 

The expression for $M$ in Eq.~(\ref{eq:PHS singularity 2}) directly manifests the fact that the degeneracy is topologically protected: 
any local smooth deformation of the time-evolution operator can only continuously change $M(t)$ via the parameters $\lambda$ and $t_0$, neither of which lifts the degeneracy.
The two possible signs of $M(t_0)$ indicate that there can in general be two types of singularities at each inversion invariant point, namely, singularities occurring at phase $0$ and at phase $\pi$.

\subsubsection*{Classification of one-dimensional systems with PHS}
\label{sec:1D PHS}
We now use the PHS-protected topological singularities described above to construct the topological classification and bulk-edge correspondence for one-dimensional systems with particle-hole symmetry.
In terms of the Floquet spectrum, a system with PHS may host topologically protected edge modes in its bulk gaps at quasienergies  
$0$ and $\pi/T$. 
The parities $\nu_{\rm PH}(0)$ and $\nu_{\rm PH}(\pi/T)$ of the numbers of edge modes in these two Floquet gaps are topological invariants.
In Ref.~\onlinecite{1D_Majorana}, 
$\nu_{\rm PH}(0)$ and $\nu_{\rm PH}(\pi/T)$ were identified as the numbers of times the phase-bands at inversion-invariant momenta cross phase $0$ and $\pi$, respectively. 
We now use the framework developed above to 
develop a more complete understanding of this relationship and to explicitly prove it.

As for the two-dimensional systems discussed in Sec.~\ref{sec:2dCase}a, the phase bands of the 1D particle-hole symmetric system can be freely deformed (while maintaining the symmetry as described above) anywhere except at topological singularities.  
Analogous to the procedure depicted in Fig.~\ref{fig:NSC:Singularity_graph}, we deform the phases $\phi_n$ for $n = 1, \ldots, N$ to zero everywhere, except around the phase-$\pi$-singularities and in a short interval at the end of the driving period where we let the phases wind to their final values.
This is done in a way that keeps $U(k,T)$ fixed, and such that the conjugate bands 
follow the deformation to preserve particle-hole symmetry via relation (\ref{eq:PHS:conjugate band relation}). 
Thus it is evident that the topological classification of the one-dimensional particle-hole symmetric system should depend only on the properties of the evolution operator around any phase-$\pi$ (i.e., zone edge) singularities, and at $t=T$ (i.e., on the Floquet bands themselves).

Which characteristics of the singularities are topologically protected?
For one-dimensional systems, all (zone edge) singularities are topologically identical if the system has more than two bands:
the evolution can be deformed such that 
the vectors $|\psi_{1,2}\rangle$ in Eq.~\eqref{eq:PHS singularity} are 
the same for all singularities. 
To see this, note that any two pairs of real orthogonal vectors in a complex space of more than two dimensions can be continuously rotated into each other, if the space has dimension $3$ or more. 
%

When the vectors $|\psi_{1}\rangle$ and $|\psi_2\rangle$ are the same for all singularities, it is possible to create or annihilate pairs of singularities with opposite sign of the parameter $\lambda$ through continuous deformations of the evolution operator. 
However, since all (zone edge) singularities are topologically identical as argued above, it is possible  
through a local smooth deformation to interchange $|\psi_1\rangle$ and $|\psi_2\rangle$ for an individual singularity, thereby flipping the sign of $\lambda$.
Hence all zone edge singularities are identical, and can be created or annihilated pairwise through smooth phase-band deformations.
Thus only the {\it parities} $Q_0$ and $Q_{\pi/a}$ of the numbers of zone edge singularities at $k = 0$ and $k = \pi/a$, respectively, are topologically invariant (here $a$ is the lattice constant). 
The special case of two bands is discussed at the end of the section.

In addition to describing the singularities, we must also keep track of the invariants of the Floquet bands, which are encoded in the 
effective Hamiltonian 
\be 
\Heff(k)=\frac{1}{T} \sum_n \left[\phi_n (k,T) P_n(k,T) +\phi_{\bar n}(k,T)P_{\bar n}(k,T)\right].
\notag \ee
The Floquet bands are completely characterized by the two standard invariants $\eta_0$ and $\eta_\pi$ 
for a non-driven particle-hole symmetric system (see, e.g.,~Ref.~\onlinecite{Periodic_Table_2}), where $\eta_k = \sgn\, \Pf\, [\Heff(k)]$ for $k=0,\pi /a$. 

Collecting the invariants identified above, we thus find that a generic one-dimensional (translationally-invariant) driven system with particle-hole symmetry is  fully characterized by the \textit{four} $\mathbb{Z}_2$ invariants
\be 
\left(\eta_0, \eta_{\pi/a},(-1)^{Q_0}, (-1)^{Q_{\pi/a}} \right).
\label{eq:PHSInvariants} 
\ee
Compared to the case without driving, translationally invariant periodically-driven systems with particle-hole symmetry are characterized by two additional $\mathbb{Z}_2$ invariants. 
The additional invariants relate to phase band singularities in the zone-edge gap (i.e., at phase $\pi$).

We now seek a bulk-edge correspondence which gives the edge mode parities $\nu_{\rm PH}(0)$ and $\nu_{\rm PH}(\pi/T)$ for the gaps at quasienergies 0 and $\pi/T$ in terms of the four numbers in Eq.~(\ref{eq:PHSInvariants}). 
To see which bulk invariants determine the edge mode parity $\nu_{\rm PH}(\pi/T)$ in the quasienergy gap at $\pi/T$, we first note that the bulk evolution can be smoothly deformed to ``flatten'' the Floquet bands to zero quasienergy.
That is, we may continuously transform $U(T) \rightarrow 1$ (or, equivalently, $H_{\rm eff} \rightarrow 0$), without closing the gap at $\pi/T$.
Such a deformation cannot change the parity of the number of edge modes in the open gap, and therefore $\nu_{\rm PH}(\pi/T)$ should not depend on $\eta_0$ or $\eta_{\pi/a}$. 
Furthermore, note that a real-space gauge transformation can be performed on the 1D system to make $k\rightarrow k+\pi/a$.
However, a gauge transformation cannot change the number of edge modes $\nu_{\rm PH}(\pi/T)$ which appear in the gap when an edge is created.
Therefore $Q_0$ and $Q_{\pi/a}$ should appear symmetrically in the final expression.
Thus we are led to the expression: $\nu_{\rm PH}(\pi/T)=\pm(-1)^{Q_0+Q_{\pi/a}}$. 
 Considering the special case $H(k,t)=0$
fixes the sign, giving
\be 
\nu_{\rm PH}(\pi/T)=(-1)^{Q_0+Q_{\pi/a}},
\label{eq:PHS:QePiInvariants}
\ee
where $\nu_{\rm PH}(\pi/T)=-1$ corresponds to an odd number of Floquet-Majorana edge modes with quasienergy $\pi/T$.

For the gap at quasienergy zero, we 
identify which combinations of the numbers in Eq.~\eqref{eq:PHSInvariants} are invariant when we allow the gap at quasienergy $\pi/T$ to close.
By closing this gap, we can remove the zone-edge singularities without changing the number of edge modes at quasienergy $0$. 
However, each time we remove a singularity at crystal-momentum $k$ in this way, $\eta_k$ changes its sign 
(see Appendix~\ref{app:SignChange}). 
Hence the only quantities that are invariant when we allow the gap at quasienergy $\pi/T$ to close are the numbers $(-1)^{Q_k}\eta_k$ for $k=0,\pi/a$.
From similar arguments as above, the expression should be symmetric in $k=0$ and $k=\pi/a$, and we conclude
\be 
\nu_{\rm PH}(0)=\eta_0 \eta_{\pi/a} (-1)^{Q_0+Q_{\pi/a}},
\label{eq:PHS:QeZeroInvariants}
\ee
where $Q_0$ and $Q_{\pi/a}$ were the numbers of zone-edge singularities at crystal momentum $0$ and $\pi/a$, respectively.  
The results in Eqs.~\eqref{eq:PHS:QePiInvariants}~and~\eqref{eq:PHS:QeZeroInvariants} agree with those obtained in Ref.~\onlinecite{1D_Majorana}. 
We see that a system can have a non-trivial edge mode spectrum if the evolution contains an odd number of zone-edge singularities,  even if the bulk Floquet operator is trivial (i.e., if $\eta_0 = \eta_{\pi/a} = 0$).

%

We finally briefly discuss the special case of two bands. 
For two-band systems, it is always possible to choose $|\psi_1\rangle=(1,0)$ and $|\psi_2\rangle =(0,1)$ in Eq.~\eqref{eq:PHS singularity}. 
Then, the sign of the parameter $\lambda$ in Eq.~\eqref{eq:PHS singularity 2} is forbidden to change, and defines a conserved charge for each singularity.
The net charges of zone-edge  singularities at $k=0$ and $k=\pi/a$ define two  topologically invariant $\mathbb Z$ indices of the system. 
This richer classification for two-band systems relies on unbroken translational symmetry, however, and we do not expect it to survive disorder. 

\subsection{
Time-reversal symmetry  with $S^2=-1$}
\label{sec:TR}

\begin{figure}\begin{center}
\includegraphics[scale=0.58]{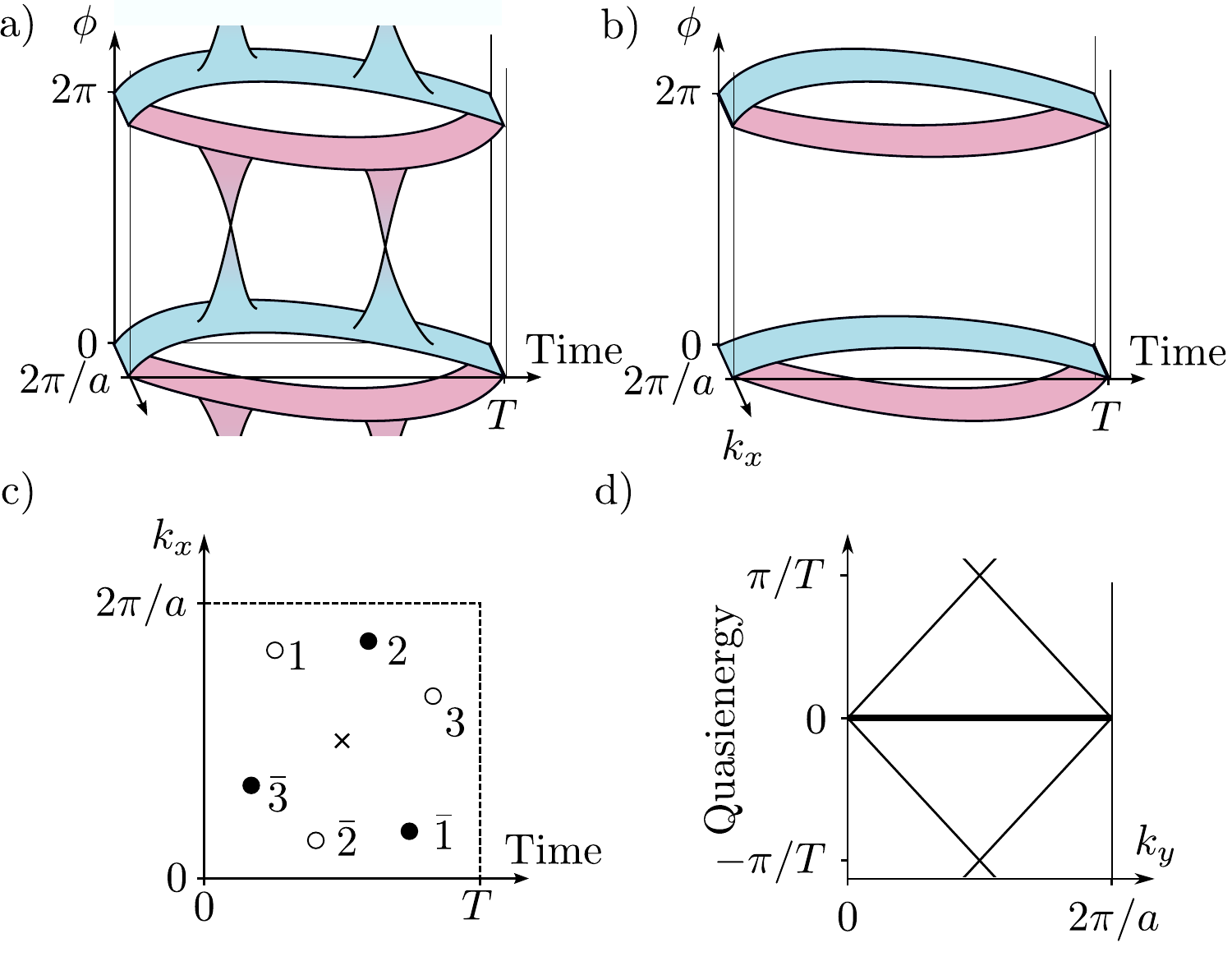}
\caption{Phase bands and topological singularities for systems with time-reversal symmetry, Eq.~(\ref{eq:TR_BareSymmetryCondition}).
After deforming to a {\it time-periodic} evolution $\tilde{U}(\vec{k},t)$ 
which has its zone-edge gap open,
two distinct classes of evolutions are possible, with an even or odd number of pairs of zone-edge topological singularities. 
a) Phase bands of $\tilde U(t)$ for a system with an odd number of singularities, and hence one pair of protected helical edge modes its zone-edge gap.
b) Phase bands of $\tilde U(t)$ for a trivial system, yielding no edge modes in its zone-edge gap. 
c) Example distribution of zone-edge singularities, with their locations in the $k_x,t$-plane depicted.
The charges are indicated by the filling of the dots, and bars indicate conjugate partners. 
d) Quasienergy band-structure of the model \eqref{eq:HTRI}, when defined in a strip geometry along the $y$-direction. 
Here we only display the edge spectrum for one edge. 
}


\label{fig:TR band structure}
\label{fig:2 band TR model}
\end{center}\end{figure}


As a final application, we now apply our framework to periodically driven systems with time-reversal symmetry, identified as a time non-local symmetry of the form in Eq.~\eqref{eq:timenonlocal} where the symmetry operator $S$ is anti-unitary and squares to $-1$ (analogous to AZ class AII).
The condition $S^2 = -1$ implies that the system must have an {\it even} number of bands, taken to be $2N$ in the discussion below.
Following the discussion in Sec.~\ref{sec:Symmetries intro}, the presence of this symmetry implies that a basis and time-origin exist such that the Hamiltonian of the system satisfies $H(\kv,t)=\sigma H^*(-\kv,T-t)\sigma$,
 where $\sigma$ is a unitary, Hermitian matrix that is purely imaginary in the specified basis (see also, e.g., Ref.~\onlinecite{2D_TR}).
In this basis, the evolution operator $U(\kv,t)$ satisfies 
\be 
U(\kv,t) = \sigma U(-\kv,T-t)^*U^{\rm T}(-\kv,T) \sigma.
 \label{eq:TR_BareSymmetryCondition}
 \ee 
Note that by substituting $t = T$, and using $U(\kv, 0) = 1$, we recover the time reversal symmetry condition on the Floquet operator itself: $U(\kv, T) = \sigma U^{\rm T}(-\kv, T)\sigma$.

We begin  our discussion of time-reversal invariant (TRI) systems below by constructing the general bulk-edge correspondence for two-dimensional TRI periodically driven systems. 
In doing so, we find that the edge mode parities in the bulk gaps can be non-trivial even if the system has trivial Floquet bands (see also Ref.~\onlinecite{2D_TR}). 
This allows driven systems to have edge mode spectra that are impossible to obtain in non-driven systems.
In particular, as we demonstrate subsequently in an illustrative example, a driven two-band system can have helical edge modes in its  Floquet zone-edge gap.
This behaviour is in contrast to that of a non-driven two band system, where Kramers' theorem  guarantees a gapless bulk.
As for the cases with no symmetries or with particle hole symmetry, 
we find a close connection between these anomalous edge mode phenomena and the appearance of topological singularities in the zone-edge gap of the bulk evolution.
After working through the example, we conclude with the topological classification for three-dimensional TRI systems.

\subsubsection*{Bulk-edge correspondence for TRI systems in two dimensions}
To simplify the derivation of the bulk-edge correspondence for two-dimensional (2D) TRI systems, we start by highlighting some general properties of the Floquet bands of such systems. 
For a periodically-driven system with time reversal symmetry in any dimension, Floquet bands $2m-1$ and $2m$ are related by time-reversal symmetry and the gap between the bands closes at the inversion-invariant points in the Brillouin zone (in accordance with Kramers' theorem). 
This holds for all $m$, and only ``even'' gaps $2m$ may thus remain open. 
Here we use the specific assignment of even and odd indices defined by the ordering scheme of Sec.~\ref{sec:NaturalQuasiEnergyZone}.

In a finite geometry, a 2D TRI system can have protected edge modes in its bulk gaps. 
Time reversal symmetry requires any chiral modes to come in time-reversal conjugate pairs; Kramers' theorem guarantees that an {\it odd} number of such pairs cannot be gapped out by any time reversal symmetry preserving perturbation.
The parity $\nu_{\rm TR}(m)$ of the number of such ``helical'' edge mode pairs appearing in gap $2m$ is thus a topological invariant that can be associated with gap $2m$. 

We now set out to find an expression for $\nu_{\rm TR}(m)$. 
First, noting that for each $m$ the Floquet bands $2m$ and $2m - 1$ are related by time-reversal symmetry, to each such pair we associate a Fu-Kane $\mathbb{Z}_2$ index $z_m$, just as for the bands of a TRI non-driven system\cite{Pfaffian_Index}.
Spectral flow arguments\cite{Pfaffian_Index}, which must hold for both static and Floquet bands, show that the relative edge mode parities of
gaps $2n$ and $2n-2$ (i.e., above and below the pair of time-reversal conjugate bands $2n$ and $2n-1$) are captured by $z_n$. 
For a system with $2N$ bands, we can thus write
\be 
\nu_{\rm TR}(m)=\nu_{\rm TR}(N) \prod_{n=1}^m z_n,
\label{eq:numN}
\ee
where $\nu_{\rm TR}(N)$ is the edge mode parity in the zone-edge gap. 
Note that $\prod_{n=1}^Nz_n$ must always be unity; the prefactor $\nu_{\rm TR}(N)$ ensures that the correct edge mode parity is recovered when setting $m = N$ in Eq.~(\ref{eq:numN}). 

To find 
$\nu_{\rm TR}(N)$ 
we first simplify the symmetry condition in Eq.~\eqref{eq:TR_BareSymmetryCondition} by smoothly deforming $U(\kv,t)$ into an evolution 
whose Floquet operator is the identity.
During the deformation we preserve TRI and keep the zone-edge gap open, such that the edge mode parity $\nu_{\rm TR}(N)$ in the zone-edge gap remains unchanged. 
%
An example of such a deformation is given in Ref.~\onlinecite{2D_TR}.
We first define an effective Hamiltonian 
$ H_\textrm{eff}(\kv)=\sum_n P_n(\kv,T)\varepsilon_n(\kv),$
where $P_n(\kv,T)$ is the Floquet eigenstate of band $n$, and $\varepsilon_n (\kv)= \phi_n(\kv,T)/T$ is the corresponding quasi-energy of the band 
(see Sec.~\ref{sec:NaturalQuasiEnergyZone} for the quasienergy zone convention). 
The family of evolutions $U(\kv,t;\alpha)=U(\kv,t)e^{i\alpha H_{\rm eff}(\kv) t}$ for $\alpha\in[0,1]$ defines a smooth interpolation of evolutions from $U$ to a time-periodic evolution $\tilde U$, given by 
\be 
\tilde U(\kv,t)=U(\kv,t)e^{iH_{\rm eff}(\kv) t}.
\ee
The Floquet operator associated with the evolution $U(\kv,t;\alpha)$ has quasi-energies $\{(1-\alpha)\varepsilon_n(\kv,T)/T\}$; the spectrum uniformly contracts, and thus the zone-edge gap stays open throughout the interpolation\cite{footnote:zone-edge}.
Given that TRI is also preserved during the deformation, $\nu_{\rm TR}(N)$ must be the same for $U$ and $\tilde U$. 

Using the fact that $\tilde U(\kv,T)=1$, symmetry condition~\eqref{eq:TR_BareSymmetryCondition} simplifies to $\tilde U(\kv,t)=\sigma \tilde U^{*}(-\kv,T-t)\sigma$. 
An evolution operator with this property can be written as 
\be
\tilde U(\kv,t)=\sum_{n=1}^N \left[P_n(\kv,t)e^{-i\phi_n(\kv,t)} + P_{\bar n}(\kv,t)e^{-i\phi_{\bar n}(\kv,t)} \right],
\label{eq:TR_Evolution}
\ee
where  the bands $\{P_n,P_{\bar n}\}$ for $n=1\ldots N$ are all  orthogonal. 
We give a specific prescription for defining the labels $n$ and $\bar{n}$ below. 

In contrast to the particle-hole symmetric case, TR symmetry relates conjugate bands at different times:
\begin{eqnarray}
P_{\bar n}(\kv,t)&=&\sigma P^{\rm T}_n(-\kv,T-t) \sigma 
\label{eq:TR_Band_relation1}\\
\phi_{\bar n}(\kv,t)
&=&-\phi_n(-\kv,T-t).
\label{eq:TR_Band_relation2}
\end{eqnarray}
For the last equality we used that $\phi_n(\kv,T)=0$.
We assign labels to the bands such that all $\phi_n(\kv,T/2)$ are positive, and then order these bands according to increasing $\phi_n$.
Their conjugate partners then follow from Eq.~(\ref{eq:TR_Band_relation2}).
In Figs.~\ref{fig:2 band TR model}a,b we show two examples of time-periodic phase-band structures that 
satisfy the time reversal symmetry above.

We now smoothly deform one half  of the phase bands of $\tilde U$, via the phases $\{\phi_n(\kv,t)\}$, in a way that preserves the boundary condition $\phi_n(\kv,T)=0$, while keeping the projectors $\{P_n(\vec{k},t)\}$ constant.
The other half of the bands follow in accordance with the symmetry above.
Through considerations similar to those made in Sec.~\ref{sec:TOP_singularities}, we find that the phase bands of $\tilde U$ can 
be deformed to zero everywhere in $\kv,t$-space except around its zone-edge topological singularities (which connect bands $N$ and $\bar N$). 
The edge mode parity $\nu_{\rm TR}(N)$ is thus completely determined by the constellation of zone-edge singularities of $\tilde U$.

Next we consider which features of the zone-edge singularity 
constellation are topologically invariant. 
As for the case without symmetries, through smooth deformations of $\tilde U$ we can annihilate singularities of opposite charges.
If $\tilde U$ has a zone-edge singularity with charge $q$ at $(\kv,t)$, symmetry dictates that it has another with charge $-q$ at $(-\kv,T-t)$; see 
Fig.\ref{fig:TR band structure}c for an example distribution of zone-edge singularities. 

In general, when two singularities annihilate, their conjugate partners must annihilate as well. 
Importantly, 
 conjugate singularities cannot annihilate directly with each other (see Appendix~\ref{APP:TRSingularities}). 
Therefore it is only possible to annihilate {\it two singularity pairs} at a time.
To give a concrete example, if singularities $1$ and $2$ in Fig.~\ref{fig:TR band structure}c annihilate each other, then by symmetry singularities $\bar 1$ and $\bar 2$ will annihilate as well. 
Singularities $3$ and $\bar 3$ will then remain, with no way to be eliminated without closing the zone edge quasienergy gap.

According to the arguments above, the parity $p$ of the number of singularity pairs is 
invariant under any smooth deformation of the evolution 
that keeps the zone-edge gap open.
Conversely, any two evolutions $\tilde U$, $\tilde U'$ with the same parity $p$ can be deformed into each other.
In particular, if $\tilde U$ has an even number of singularity pairs, it can be smoothly deformed into the identity.
Evolutions $\tilde U$  therefore fall into two classes: those with an odd number of zone edge singularity pairs (Fig.~\ref{fig:2 band TR model}a) and those with an even number of pairs (Fig.~\ref{fig:2 band TR model}b).
Evolutions within the same class must have the same edge mode parity $\nu_{\rm TR}(N)$, since they can all be related by smooth deformations that preserve $\nu_{\rm TR}(N)$.
Thus $\nu_{\rm TR}(N)$ must depend only on $(-1)^{p}$. 
Since we already found that evolutions with even $p$ are topologically trivial (i.e., smoothly deformable to the identity), we identify
\be
\nu_{\rm TR}(N)=(-1)^p.
\label{eq:zoneedgeparity}
\ee
The edge mode parity in the zone-edge gap is thus given by the parity of the number of singularity pairs of $\tilde U$. 
%
%
%
%
%
%
%

Combining results (\ref{eq:numN}) and (\ref{eq:zoneedgeparity}), we then obtain for the edge mode parity of gap $m$:
\be 
\nu_{\rm TR}(m)=(-1)^{p}\prod_{n=1}^m z_n,
\label{eq:TR_BulkEdge}
\ee
where $p$ is the parity of the number of 
zone-edge singularity pairs of $\tilde U(\kv,t)$. 
We see that even if the bulk Floquet operator of a system is trivial (i.e.,~$z_n=1$ for all $n$), the edge mode spectrum can be nontrivial.
This is precisely the case  when $\tilde U(\kv,t)$ has an odd number of singularity-pairs in its zone-edge gap (see Fig.~\ref{fig:TR band structure}ab). 

We end our discussion by noting that the ``anomalous'' $\mathbb{Z}_2$ index $\nu_{\rm TR}(N)=(-1)^p$, Eq.~(\ref{eq:zoneedgeparity}), can be found directly as a time-domain expression in terms of $\tilde U(\kv, t)$. 
According to the time-reversal symmetry condition $\tilde U(\kv,t)=\sigma \tilde U^{*}(-\kv,T-t)\sigma$, one singularity out of each conjugate pair must occur within the first half of the driving, $0 < t < T/2$. 
Therefore $p$ is in fact equal to the
number of zone-edge singularities of $V(t)$, where $V(t)$ is the evolution given by $\tilde U(t)$ restricted to the first half of the driving\cite{footnote:T2Singularities}.
In Sec.~\ref{sec:2dCase} we found that the number of zone-edge singularities of an evolution is given by winding number of the evolution in its zone-edge gap.
Hence we have
\be 
p 
 = w_{2N}[V(t)]\ ({\rm mod}\ 2).
 \label{eq:TR_TimeDomainParity}
\ee
The above result is  consistent with the results obtained in Ref.~\onlinecite{2D_TR}.

\subsubsection*{Example: Nontrivial $\mathbb{Z}_2$ index for a two-band TRI system} 

\label{sec:2 band TR system}
To illustrate one of the new topological phenomena which arise in periodically driven systems, we now show that periodic driving allows a 2D 
TRI system with two bands to have protected helical edge modes. 
This is in contrast to the situation for non-driven systems, where a minimum of four bands is required.
We explicitly demonstrate this behavior for a specific model, 
using the results above as well as direct numerical calculation.

Consider a spin-$1/2$ particle on a square lattice with one orbital per site.
We construct the evolution based on the model with non-trivial winding numbers presented in Sec.~\ref{sec: Specific Model}, now with spin playing the role of what was an orbital index.
Specifically, we define a TRI evolution via the time-periodic Bloch Hamiltonian
\be 
\label{eq:HTRI}
H(\kv, t) = \left\{\begin{array}{ll}
H_0(\kv, 2t) , \quad &t<T/2\\
\sigma_y H_0^*(\kv, 2(T-t))\sigma_y , \quad & t>T/2
\end{array}\right.
\ee
where, as a $2 \times 2$ matrix, $H_0(\kv,t)$ has the same form as the Hamiltonian presented in Sec.~\ref{sec: Specific Model}. 
The Hamiltonian $H(\kv,t)$ is by construction time-reversal symmetric. 
In real-space, Hamiltonian (\ref{eq:HTRI}) contains on-site terms as well as nearest-neighbour and next-nearest neighbour hopping terms.  

In Ref.~\onlinecite{Winding_Number} it was noted that the Floquet operator $U_0(\kv, T)$ associated with $H_0(\kv,t)$ is unity for the parameter values $\delta_{AB}=0$, $\lambda_n=5\pi/{(2T)}$, and that 
the corresponding time evolution operator $U_{0}(\kv,t)$ has winding number $1$. 
For our model (\ref{eq:HTRI}), we choose the parameters of $H_0$ such that $\delta_{AB}=0$ and $\lambda_n=5\pi/{T}$.
The Floquet operator of the (translationally invariant) system governed by $H$ is again equal to the identity.

We now use Eq.~\eqref{eq:TR_BulkEdge} to predict the number of helical edge mode pairs, $\nu_{\rm TR}(N=1)$, that will appear for this model when defined in a strip geometry.
First, note that for the system without edges the Bloch-space evolution is periodic with its zone-edge gap open, and hence $\tilde{U}(\kv, t) = {U}(\kv, t)$.
Second, because the evolution in the first half of the driving corresponds to that of the model discussed in Sec.~\ref{sec: Specific Model}, we know that the evolution for $0 < t \le T/2$ possesses exactly one zone edge singularity.
Hence $p = 1$ and we have
\be 
\nu_{\rm TR}(1) = (-1)^p = -1. 
\ee
We thus expect one helical edge mode pair to appear in the zone edge gap, for each edge.

To confirm the analysis above, we numerically study this model in a strip geometry, with edges parallel to the $y$-direction.
The Hamiltonian of the strip geometry system  is obtained by truncating the real-space tight-binding Hamiltonian   in the $x$-direction.
Using this truncated tight-binding Hamiltonian we numerically calculate the corresponding Floquet operator, and obtain the quasienergy band structure as a function of the conserved momentum component $k_y$, see Fig.~\ref{fig:2 band TR model}d.
At each edge we find a pair of helical modes, which are time-reversal conjugates of each other.
This model thus explicitly demonstrates the existence of anomalous helical edge modes in two-dimensional periodically driven systems with TRI, and shows that non-trivial topology can be found even in a case with only two bands.

\subsubsection*{$\mathbb{Z}_2$ index for a 3D periodically driven  system with time-reversal symmetry}

The $\mathbb Z_2$ index for two-dimensional systems, see Eqs.~\eqref{eq:TR_BulkEdge}~and~\eqref{eq:TR_TimeDomainParity}, 
can also be used to define a $\mathbb Z_2$ index $\nu_{\rm 3D}(m)$ for three-dimensional (3D) systems with time-reversal symmetry.
The index $\nu_{\rm 3D}(m)$ indicates whether or not Floquet gap $m$ hosts non-trivial surface states, analogous to those of a strong topological insulator. 
In the same way as for the non-driven case, we consider the
 two 2D TRI systems defined by the restriction of $U(\kv,t)$ to the planes $k_i=0$ and $k_i=\pi/a$, where $i$ can be $x,y$ or $z$.
We then identify the index $\nu_{\rm 3D}(m)$ as the product of the indices $\nu_{\rm TR}(m)$, from Eq.~\eqref{eq:TR_BulkEdge}, for the two systems. 
Note that if the indices calculated for the planes $k_i = 0$ and $k_i = \pi/a$ 
are the same, their common index 
determines whether the system acts as 
a weak topological insulator in the plane orthogonal to $k_i$.

\section{Discussion}
\label{sec:Discussion}
In this paper we found that the ``phase-band structures'' of time evolution operators provide a powerful basis 
for visualizing and understanding the topology of Floquet-Bloch systems.
By considering smooth deformations of the phase bands, we showed that topologically protected degeneracies, or topological singularities, play a crucial role in distinguishing the topology of driven and non-driven systems. 
In particular, the presence of phase Brillouin zone edge topological singularities can present an obstruction to smoothly deforming the evolution of a driven system into one obtainable in a non-driven system.

Our work demonstrates a general method for topological classification of Floquet-Bloch systems, based on identifying all features of the phase band structure of a given system that cannot be removed by smooth deformations.
This approach appears to offer means 
to exhaustively classify Floquet-Bloch systems and to straightforwardly derive the corresponding bulk-edge correspondences.


In the cases we considered, with symmetries imposed in analogy with the ten Altland-Zirnbauer classes, we found that the edge spectra associated with individual bulk gaps of the Floquet operator have the same features as those of non-driven systems in the corresponding symmetry classes. 
However, we found in all cases that periodic driving could induce \textit{global} edge spectra that are impossible to obtain in non-driven systems.
In particular, with periodic driving, topologically protected helical edge states can be produced in time-reversal invariant systems with only two-bands, while a minimum of four bands is needed without driving.
In each case we considered, these new ``anomalous'' features were closely related with the appearance  of zone-edge singularities in the time-bulk. 
These phenomena further demonstrate that the relation between the topological properties of the bulk evolution and the appearance of protected edge modes is fundamentally changed in the driven context: the topology of a periodically driven system cannot be fully characterized by the stroboscopic Floquet operator $U(\kv, T)$ or the corresponding effective Hamiltonian alone. 

Finally, our results provide new  intuition about the topology of Floquet-Bloch systems.
While time-domain invariants such as the winding number found in Ref.~\onlinecite{Winding_Number} offer a mathematically well-defined prescription for characterizing topology in driven systems, often a clear physical picture is missing.
Here, one of our central results is that non-trivial topological phenomena in periodically driven systems appear when topological singularities are introduced into the  phase bands of the bulk  time-evolution operator.
In particular, in any driven system where the Floquet bands have different topology from that of the initial instantaneous Hamiltonian (i.e., at $t = 0$), at least one topological singularity must be encountered during the evolution.
We expect that this insight may help provide guidance for the construction of new driving protocols to realize non-trivial topological phases in periodically driven systems.

The instantaneous and time-nonlocal symmetries considered in Sec.~\ref{sec:Symmetries intro} were chosen 
to illustrate the power of the phase band framework.
The symmetries altered the topology of the bulk evolution through their ability to protect new types of topological singularities. 
However, 
we expect that other types of conditions could give rise to new non-trivial topology in the phase bands. 
The exploration of other symmetries and their role in the topology of periodically driven systems, as well as the search for good candidate systems for their realization, are interesting directions of future study.

We gratefully acknowledge E.~Berg and N.~Lindner for enlightening discussions.  
MR thanks the Villum Foundation and the People Programme (Marie Curie Actions) of the European Union’s Seventh Framework Programme (FP7/2007-2013) under REA grant agreement PIIF-GA-2013-627838 for support.

\appendix

\section{\label{app:Time-domain_Expressions_NSC} Time-domain expressions for the invariants $\nu_0$ and $\nu_1$}
In this appendix we derive a time domain integral expression for the bulk invariant $w_m[U]$ in Eq.~(\ref{eq:NuMInvariant}), which corresponds to the number of edge modes appearing in gap $m$ when the system is defined in a geometry with edges. 
We directly show that this invariant is equivalent to the winding number invariant $W[U_\varepsilon]$ of Ref.~\onlinecite{Winding_Number}, with $\varepsilon$ set equal to a quasienergy value inside gap $m$:
%
%
\begin{equation}
\label{eq:APP:NSC:WindingNumberFormula}W[U_m]=\int\!\frac{{d}^2 k {d}t}{8\pi^2} 
\Tr \left\{U_m^\dagger \dt U_m   \,[U_m^\dagger \dkx U_m , U_m^\dagger \dky U_m] \right\}. 
\end{equation}
Here $U_m$ is a time-periodic evolution operator, satisfying $U_{m}(\kv,T)=1$, which is obtained from the original evolution $U$ by a smooth deformation in which gap $m$ of the quasienergy spectrum is kept open. 
Explicitly, the time-periodic evolution operator $U_m$ can be obtained by deforming the final values of the phase bands $\phi_n(\kv,T)$ to zero for $n=1\ldots m$, while the final values of the remaining bands are deformed to $2\pi$. 

In order to demonstrate the equivalence of the two invariants, i.e., to show $w_m[U]=W[U_m]$, we first consider two special cases and then discuss the general situation. 
In this appendix we label points in the three dimensional $\kv,t$-space by a single dimensionless vector 
\be 
\vec{s}= \left(\frac{\kv}{2\pi /a},\frac{t}{T}\right).\notag
\ee


\subsection{Winding number in case of no singularities}

To begin, we first consider the case where $U_m(\kv,t)$ has no topological singularities in the zone-edge gap. 
It is then possible to continuously deform the evolution operator $U_m$ into one corresponding to a non-driven system, as described in Sec.~\ref{sec:PhaseBands}. 
In doing so, the winding number $W[U_m]$, a topological invariant, cannot change its value.
After the deformation, $W[U_m]$ is  simply the winding number of a non-driven system described by the Hamiltonian 
\be 
H_m(\kv) =\frac{2\pi}{T} \sum_{n=m+1}^N P_n(\kv,T).
\ee
The winding number of a system governed by such a Hamiltonian was found in Ref.~\onlinecite{Winding_Number} to be $\sum_{n=1}^m C_n$, where $C_n$ is the Chern number of  Floquet band $n$. 
Hence 
\be 
W\left[U_m\right]=\sum_{n=1}^m C_n, \ \textrm{(no zone-edge singularities)}.
\ee
Note that the winding number will always be zero in the  quasienergy zone-edge gap if the phase bands do not host any zone-edge singularities.
This follows from the fact that the sum of Chern numbers for all bands must evaluate to zero, $\sum_{n=1}^N C_n =0$.

\subsection{Winding number in the case of one singularity  }\label{sec:NSC:Winding number with one singularity}

\begin{figure}
\includegraphics[scale=0.58]{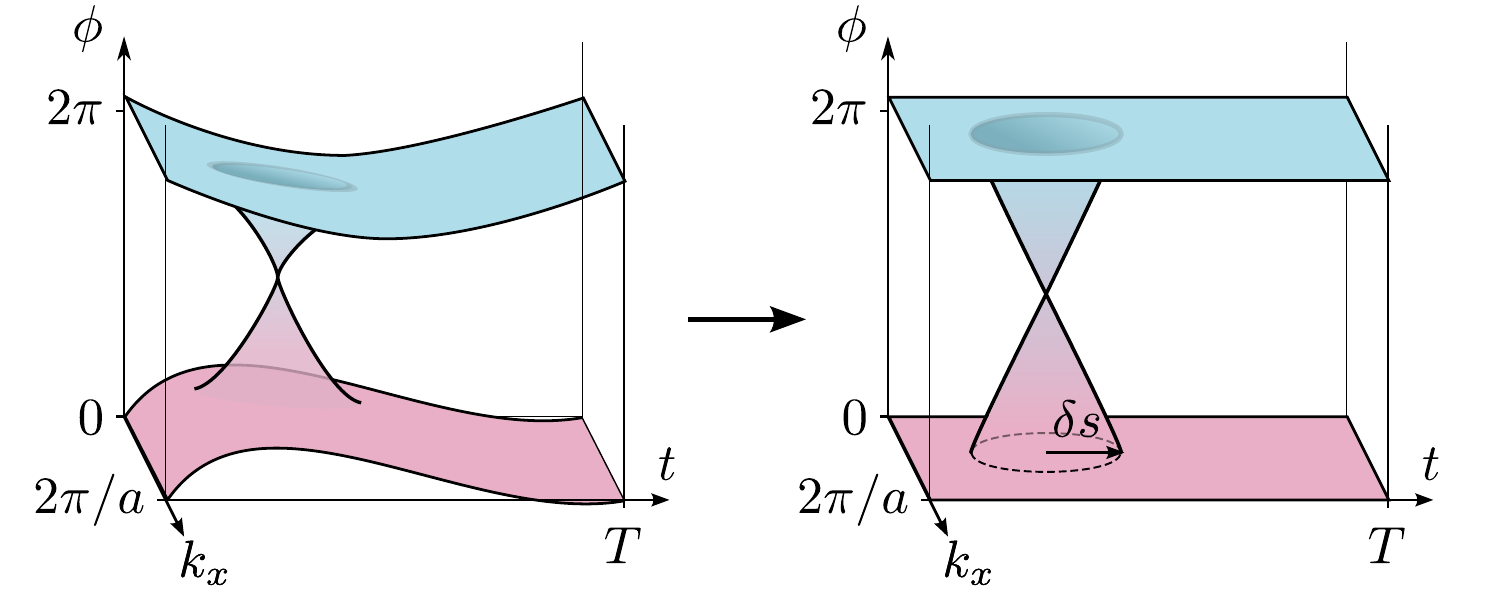}
\caption{
The deformation of an isolated singularity discussed in Sec.~\ref{sec:NSC:Winding number with one singularity}.
In the figure, phase band $1$ (blue) is shifted up by $2\pi$ for clarity of illustration. 
}
\label{fig:NSC:Singularity_Diagram}
\end{figure}

We now consider the case where $\phi_n(\kv,T)=0$ for all $n$, and  $U_m(\kv,t)$ has only one singularity in the zone-edge gap, located at 
$\vec{s}_0=(\kv_0/(2\pi /a),t_0/T)$. 
At the singularity, the two touching bands $N$ and $1$ have phases $\phi_d$ and $\phi_d-2\pi $, where  $\phi_d$ is a real number determined by details of the evolution.
In this case we can deform the phase bands to zero for all $\vec{k}, t$, except in a small spherical neighbourhood of radius $\delta s$ that surrounds the singularity
(here lengths are computed with respect to the usual norm on the dimensionless vector $\vec{s}$). 
Within the neighborhood, all $N-2$ bands not involved in the singularity can still be flattened.  
The phase values of the two intersecting bands are deformed to evolve linearly from zero at the edge of the neighbourhood to $\pi$ and $-\pi$ at the center (letting $\phi_d$ go continuously to $\pi$ in the process), see Fig.~\ref{fig:NSC:Singularity_Diagram}. 
Under the deformation we keep the eigenstates of $U_m(\vec{s})$ constant everywhere.

The deformed evolution operator $\tilde U_m$ is equal to the identity for all $\vec{s}$, except in the small region of radius $\delta s$ that surrounds the singularity (see Fig.~\ref{fig:NSC:Singularity_Diagram}). 
Within this neighbourhood, $\tilde U_m$ 
takes the form 
\be
\tilde U_m(\vec{s})=\sum_{n\neq 1,N} |\chi_n\rangle \langle \chi_n|+\!\! \sum_{a,b=1,N} |\psi_a\rangle \langle \psi_b|M_{ab}(\vec{s}),
\label{eq: deformed U}
\ee
where $M_{ab}(\vec{s})$ is a $2\times 2$ matrix whose eigenvectors are the eigenvectors of the matrix $(\vec{s}-\vec{s}_0)_jS_{jk}\sigma_k$.
Here, the real invertible $3\times 3$ matrix $S$ was defined in Sec.~\ref{sec: Topological singularities in 2D systems}, and $\{\sigma_k\}$ are the Pauli matrices.
From the description of the flattened phase bands above, 
we know that the logarithms of the eigenvalues of $M$ must grow linearly from $0$ at $|\vec{s} - \vec{s}_0|=\delta s$ to $-i\pi$ and $i\pi$ at $\vec{s}=\vec{s}_0$.  
For  $|\vec{s}-\vec{s}_0|<\delta s$, $M_{ab}$ thus takes the form
\begin{align}
M_{ab}(\vec{s})=&v^-_a(\vec{s}) v^{-*}_b(\vec{s}) e^{i \pi (|\vec{s}-\vec{s}_0|/\delta s -1)} \notag \\ &+ v^+_a(\vec{s}) v^{+*}_b(\vec{s}) e^{-i \pi(|\vec{s}-\vec{s}_0|/\delta s-1) }. 
\end{align}
 The vectors $v^-(\vec{s})$ and $v^+(\vec{s})$ are the eigenvectors of the traceless $2\times 2$ Hermitian matrix $(\vec{s}-\vec{s}_0)_jS_{jk}\sigma_k $, corresponding to negative and positive eigenvalue, respectively. 
 Since the matrix $S$ is real and invertible, we can write it as $S=R_1\Lambda R_2$, where $R_1$ and $R_2$ are orthogonal and $\Lambda$ is a diagonal matrix with positive entries (this is the singular value decomposition of $S$, see e.g.,~Ref.~\onlinecite{Singular_Value_Decomposition}). 
A continuous deformation of the entries of $\Lambda$ to $1$ results in a orthogonality-preserving continuous interpolation of the eigenvectors $v^\pm(\vec{s})$ to the eigenvectors of $R_{jk} (\vec{s}-\vec{s}_0)_j\sigma_k$, where $R=R_1R_2$.
By continuously deforming the vectors $v^\pm$ in this way,
$\tilde U_m$ is deformed into an evolution operator $V_m$ still of the form \eqref{eq: deformed U}, but with the matrix $M$ given by
\be
M(\vec{s}) = \left\{\begin{array}{ll}
1,  &|\vec{s}-\vec{s}_0|>\delta s \\
-\exp\left[ \frac{-i\pi  }{\delta s}(\vec{s}-\vec{s}_0)_i R_{ij} \sigma_j \cdot\right],  &|\vec{s}-\vec{s}_0|<\delta s.
\end{array}\right. 
\label{eq: deformed M}
\ee

Recall that $R$ is orthogonal and its determinant $|R|$ is the charge of the singularity, $q=|R_1||R_2|= |S|$, see Eq.~\eqref{eq: singularity charge definition}.  
In Appendix~\ref{sec: Winding number of singularity} we explicitly evaluate the winding number \eqref{eq:APP:NSC:WindingNumberFormula} 
of the evolution operator $V_m$.
We find:
\be 
W[V_m]=-|R| \equiv -q.
\ee
Using the fact that the winding number could not change during the deformation from $U_m$ to $V_m$, we thus establish 
\be 
W[U_m]= -q .
\label{eq: Winding number with one singularity}
\ee
In other words,  if $U_m$ contains one isolated singularity, the winding number of $U_m$ is given by the corresponding 
 charge of the singularity (with a minus sign).  


\subsection{The general case}
We now consider the general case, where $U_m$ has $N$ topological singularities in the zone-edge gap, with charges $\{q^{(m)}_i\}$.
In order to evaluate the winding number, we deform $U_m(\kv,t)$ as described in Sec.~\ref{sec:NSC:Complete_classification} and shown in Fig.~\ref{fig:NSC:Singularity_graph}.
 The deformed evolution is the identity everywhere except for in small isolated regions surrounding the  singularities  (region I), as well as in the short ramping time-interval $\delta t$ at the end of the driving (region II).

The winding number \eqref{eq:APP:NSC:WindingNumberFormula}
is defined as 
an integral over $\kv,t$-space of the quantity  $F_m(\kv,t) =\frac 1 {8\pi^2} \Tr \{U^\dagger_m \dt U_m \, [U^\dagger_m \dkx U_m,\,   U^\dagger_m \dky U_m] \}$. 
For the deformed, ``band-flattened,'' system, $F_m$ is  only nonzero in each of the  isolated regions that surround the singularities, and in the final ramp region II. 
We can therefore split up the integral of $F_m$  into a sum of integrals over each of these nontrivial regions. 

From the first special case we examined, i.e., for an evolution with no singularities,  we know that the integral of $F_m(\kv,t)$ over  region II equals $\sum_{n=1}^m C_n$, where $\{C_n\}$ are the Chern numbers of Floquet bands $1 \ldots m$. 
From the second special case, we know that the integral of $F_m(\kv,t)$ over one of the regions surrounding a zone-edge  singularity equals minus the  charge of the singularity, i.e.,  $-q$.

Summing the integrals over all regions, we obtain:
\be 
W[U_m]=\sum_{n=1}^m C_n - \sum_{i=1}^N q^{(m)}_i.
\ee 
Note that $U_m$ can be 
constructed  by deforming the phase bands of $U$ only at the end of the driving. 
Therefore the net charge of all zone-edge singularities in the time-bulk should be the same for $U$ and $U_m$. 
Thus $\sum_i q_i^{(m)}=\sum_i q_i^{\rm (ZES)}$, where $\{q_i^{\rm (ZES)}\}$ are the zone-edge singularity charges for the original system with evolution governed by $U$. 
Hence, we finally have the result for the number of edge modes in a two-dimensional system:
\be 
W[U_m]=\sum_{n=1}^m C_n -\sum_i q_i^{\rm (ZES)}  = w_m[U].
\ee
This is what we set out to show.

\section{Derivation of Eq.~\eqref{eq: Winding number with one singularity}}
\label{sec: Winding number of singularity}

In this appendix we prove that the winding number \eqref{eq:APP:NSC:WindingNumberFormula} of an evolution operator 
$V_m$ of the form in Eq.~\eqref{eq: deformed U}, with the matrix $M$ given in \eqref{eq: deformed M},  is equal to $-|R|$.
We begin by inserting $V_m$ from Eqs.~\eqref{eq: deformed U} and (\ref{eq: deformed M}) into Eq.~(\ref{eq:APP:NSC:WindingNumberFormula}), to obtain
\begin{align} 
W=&\frac{\epsilon_{ijk}}{24 \pi^2} \int_{|\vec{s}-\vec{s}_0|< \delta s}\!\!\!\!\!\!\!\!\!\!\!\!\!\!\!\!\!  d^3 \vec{s}\,    \Tr \left\{ M^\dagger \partial_{\vec s_i}  M \,  M^\dagger \partial _{\vec s_j}  M \,  M^\dagger \partial _{\vec s_k}  M\right\},
\end{align}
where $\varepsilon_{ijk}$ is the Levi-Civita symbol. 
Summation over repeated indices  is used and will be used in the rest of this appendix.

In order to exploit the $\vec s$-space 
spherical symmetry of the deformed evolution $V_m$, we shift from Cartesian coordinates to spherical coordinates centred around $\vec{s}_0$, defined such that
\be 
\vec{s}-\vec{s}_0\ \equiv\ (s\, \sin\theta\sin\phi,\,s\, \sin\theta\cos\phi,\,s\,\cos\theta). 
\ee
After the coordinate transformation, $W$ is expressed as
%
%
\begin{align}
W= \frac{\epsilon_{ijk} }{24 \pi^2}&\int _0^{\delta s} \!{d}s \int _0 ^\pi  \!{d}\theta \int_0^{2\pi} \!{d}\phi\ |J|  J_{i\alpha}J_{j\beta} J_{k\gamma}  
\notag \\ &\cdot \Tr \left\{ M^\dagger \partial_\alpha  M \,  M^\dagger \partial _\beta  M \,  M^\dagger \partial _\gamma M\right\},
\end{align}
where 
$J$ is the Jacobian matrix of the coordinate transformation, and the Greek letters $\alpha,\beta,\gamma$ run over the coordinates $s,\theta,\phi$.
We now use the following useful identity for the Levi-Civita symbol that holds for  any real invertible $3\times 3$ matrix $A$ \cite{Levi-Civita}:
\be 
A_{i\alpha}A_{j\beta} A_{k\gamma} \epsilon_{ijk} = \frac{\epsilon_{\alpha\beta\gamma}}{|A|}.
\label{eq: Levi Civita identity}\ee 
%
%
With the help of this identity we see that the Jacobian matrices always cancel out:
\begin{align*}
W=&\frac{\epsilon_{\alpha \beta \gamma}}{24 \pi^2}\int _0^{\delta s}  \!{d}s \int _0 ^\pi  \!{d}\theta \int_0^{2\pi} \! {d}\phi \\
&  \Tr \left\{ M^\dagger \partial_\alpha  M \,  M^\dagger \partial _\beta  M \,  M^\dagger \partial _\gamma M\right\}.
\end{align*}
Summing over the indices, we obtain:
\begin{align*}
W=\frac{1}{8 \pi^2}&\int _0^{\delta s}  \!{d}s \int _0 ^\pi  \!{d}\theta \int_0^{2\pi} \! {d}\phi \\
&  \Tr \left\{ M^\dagger \partial_s  M \,  [M^\dagger \partial _\theta  M ,\,  M^\dagger \partial _\phi M]\right\}. 
\end{align*}
Using the cyclic property of the trace as well as the identity $\partial M M^\dagger = - M \partial M^\dagger$, we get
\begin{align}
W=-\frac{1}{8\pi^2}&\int _0^{\delta s}\! {d}s \int _0 ^\pi\! {d}\theta \int_0^{2\pi}\!  {d}\phi  \notag \\
&\Tr \left\{ M^\dagger \partial_s  M  \, [\partial _\theta M^\dagger ,\, \partial _\phi M]\right\}. 
\end{align}
We now consider the explicit canonical form of $M$, Eq.~(\ref{eq: deformed M}), in polar coordinates, in the region $|\vec{s}-\vec{s}_0|\le \delta s$:
\be 
M(\vec{s},\theta,\phi)=- \exp\left( -\frac{i\pi  }{\delta s}(\vec{s}-\vec{s}_0) \cdot \tauv\right),
\ee
where $\tau_i = R_{ij}\sigma_j$. 
Defining $\hat s (\theta,\phi)= (\vec{s}-\vec{s}_0)/s$, we evaluate each of the factors in the integrand 
\begin{align*}
M^\dagger \partial _s M &=
 -\frac{i \pi }{\delta s} \ \hat s \cdot \tauv 
\\
\partial_{\theta} M &= i\sin\left(\frac{ \pi s}{\delta s} \right)\partial_{\theta} \hat s \cdot \tauv \\ 
\partial_{\phi} M &= i\sin\left(\frac{ \pi  s}{\delta s} \right)\partial_{\phi} \hat s \cdot \tauv .
\end{align*}
Hence, after performing the integral over $s$, we obtain
\begin{align}
W= \frac {i   }{16\pi } \int _0 ^\pi\!  d\theta \int_0^{2\pi}\!  d\phi\,  
\Tr \left\{ \hat s \cdot \tauv    \, [ \partial _\theta \hat s \cdot \tauv   ,\,  \partial _\phi \hat s \cdot \tauv ]  \right\}. 
\label{eq: final integral}
\end{align}
Working on the integrand, we note:
\begin{align}
&\Tr \left\{ (\hat s \!\cdot\! \tauv) (\partial _\theta \hat s \!\cdot\! \tauv)  ( \partial _\phi \hat s \!\cdot\! \tauv)  \right\}  
  \notag \\&\quad\quad\quad\quad\ \ =
\hat s_i\, \partial_\theta \hat s_j\, \partial_\phi \hat s_k \,R_{ia}R_{jb}R_{kc}\,\Tr \left\{\sigma_a \sigma_b \sigma_c\right\} .
\end{align}
%
Using the Pauli matrix identity $\Tr \{ \sigma_i \sigma_j \sigma_k\} = 2i\epsilon_{ijk}$, we obtain  
\begin{align*}
\Tr \left\{ (\hat s \!\cdot\! \tauv)  (\partial _\theta \hat s \!\cdot\! \tauv)  (\partial _\phi \hat s \!\cdot\! \tauv)  \right\}  
&= \hat s_i \partial_\theta \hat s_j \partial_\phi \hat s_k \cdot 2iR_{ia}R_{jb}R_{kc}\epsilon_{abc}  \\
&=\hat s_i \partial_\theta \hat s_j \partial_\phi \hat s_k \cdot 2i|R|\epsilon_{ijk}.
\end{align*}
Here we used the Levi-Civita symbol identity \eqref{eq: Levi Civita identity}, and the fact that $|R|^{-1}=|R|$. 
Restoring antisymmetry in $\theta$ and $\phi$, and going back to vector notation, we have 
\begin{align}
\Tr \left\{ (\hat s \!\cdot\! \tauv) [(\partial _\theta \hat s \!\cdot\! \tauv),\, (\partial _\phi \hat s \!\cdot\! \tauv)]  \right\} 
&= 
4i |R|\  \hat s \cdot (\partial_\theta \hat s \times \partial_\phi \hat s)  \notag \\
&= 4i |R| \sin \theta .
 \end{align}
 Hence the integrand in Eq.~\eqref{eq: final integral} is simply $4i|R|$ times the surface area element of the sphere. We thus have 
 \begin{align}
W &= \frac {i} {16\pi} \int _0^\pi\!\textrm{d}\theta \int _0 ^{2\pi}\!\textrm{d}\phi\ 4i|R|\sin \theta   \notag 
 \\
 &= - |R |\notag 
 \end{align}
Recalling that $q=\sgn|S|=|R|$, we see that the winding number contribution of an isolated singularity is given by the charge $q$ of the singularity. 

\section{ Sign change of the Pfaffian for zone-edge singularities  passing through $t = T$ }
\label{app:SignChange}
In Sec.~\ref{sec:1D PHS} of the main text, on the topological classification of one-dimensional systems with particle hole symmetry, we made use of the fact that the sign of $\Pf\,[H_\textrm{eff}]$ switches when a phase-$\pi$ (i.e., zone edge) singularity passes through the plane $t=T$ due to a continuous deformation of the evolution.
In this appendix we give a proof of this statement.

We begin by introducing an alternative labeling for the phase bands at the inversion-invariant points in the Brillouin zone. 
A crucial point in this paper is that it is not always possible to define the eigenstate projectors $\{P_n(k,t)\}$ of the evolution operator such that they are continuous everywhere in $(k,t)$-space.
However, we can always define them such that they are continuous along an arbitrary {\it line} in $k,t$-space. 
At an inversion-invariant point $k_{\rm inv}$ in the momentum Brillouin zone, we can therefore write the evolution operator of a particle-hole symmetric system in the form
\newcommand{\kkinv}{k_\textrm{inv}}

\be 
U(\kkinv,t)= \sum_n\left[\hat P_n(t)e^{-i\hat \phi_n(t)}-\hat P_n^{\rm T}(t) e^{i\hat \phi_n(t)}\right], 
\ee
where both the projectors $\hat P_n(t)$ and the phases $\hat \phi_n(t)$ are continuous functions of $t$, and we suppress the momentum index $k_{\rm inv}$ for brevity. 
Here, the projector  $\hat P_n^{\rm T}$ is the transpose of $\hat P_n$ in the basis where $H(k,t)=-H^*(-k,t)$. 

When we impose the 
requirement that the projectors $\{\hat P_n(t)\}$ are continuous in $t$, the labeling scheme defined in Sec.~\ref{sec:PHS} cannot be employed. 
In particular, 
in the original scheme the phases $\{\phi_n(k,t)\}$ were constrained to live within the interval $[0, \pi]$, 
``bouncing'' off the phase Brillouin zone edge when meeting a conjugate partner band there.
Under the new construction, continuity of the projectors requires that the phases  of conjugate bands pass through each other, 
such that a phase $\hat \phi_n(t)$ may move from inside to outside the interval $[0,\pi]$. 
We note that such crossings of partner bands are protected by symmetry and can occur at phases which are either even or odd multiples of $\pi$.
We identify them respectively as phase-$0$ and phase-$\pi$ topological singularities.

We now use this 
picture of topological singularities
 to show that the Pfaffian of $H_{\rm eff}$ changes its sign when a phase-$\pi$ singularity passes through the plane $t=T$.
We first express the effective Hamiltonian at the inversion invariant point $\kkinv$, $H_\textrm{eff}(\kkinv)$, in terms of the continuously defined phase bands $\hat \phi_n$ and $\hat P_n$:
\be 
H_\textrm{eff}(\kkinv)=\sum_n F\big(\hat \phi_n(T) \big) \left[\hat P_n(T)-\hat P_n^\mathrm{T}(T)\right]. 
\ee
Here $F$ is the modulo $2\pi$ function, defined such that it takes value between $-\pi$ and $\pi$. 
The Pfaffian of $H_{\rm eff}$ can then be written as (see, e.g., Ref.~\onlinecite{Pfaffian}):
\be 
\Pf\,[H_\textrm{eff}(\kkinv)]=\Pf\left[\sum_n (\hat P_n(T)-\hat P_n^\mathrm{T}(T))\right]\prod_n F\big(\hat \phi_n(T)\big) .\label{eq:Pfaffexp}
\ee
Above 
we identified phase-$\pi$ singularities as crossings of partner bands $\hat \phi_n(t)$, $\hat \phi_{\bar n}(t)$ through the phase Brillouin zone edge at phase $\pi$. 
Hence, under a continuous deformation where a phase-$\pi$ singularity passes through the plane $t=T$, the value of $\hat \phi_n(T)$ for some $n$ crosses through an odd multiple of $\pi$. 
As $\hat{\phi}_n(T)$ passes through the quasienergy zone edge, $F(\hat \phi_n(T))$ therefore changes it sign.
During this process, $F(\hat \phi_n(T))$ for all other $n$ and {\it all} of the eigenstate projectors $\{\hat P_n\}$ may only change infinitesimally. 
From the above expression for the Pfaffian, Eq.~(\ref{eq:Pfaffexp}), it then follows that $\Pf\,[H_\textrm{eff}(\kkinv)]$ changes its sign when a phase-$\pi$ singularity passes through $t=T$.

\section{Proof that TR-conjugate singularities cannot annihilate}
\label{APP:TRSingularities}
%
In this appendix we show that two TR-conjugate zone-edge singularities cannot annihilate with each other,  as claimed in Sec.~\ref{sec:TR}.
The argument 
follows  indirectly  from the main text.

The discussion in Sec.~\ref{sec:TR} implies that if it were possible for conjugate singularities to annihilate, the evolution of any system with TRI could  be smoothly deformed to the identity, while keeping the zone-edge gap open. 
However, in Sec.~\ref{sec:2 band TR system} we saw an example of a system with one pair of helical edge modes in its zone-edge gap.
Suppose it were possible to deform the evolution of this system into the identity, without breaking TRI or closing the zone-edge gap.
Then the helical edge modes would have to disappear without a quasienergy gap closing. 
However, this is impossible and implies a contradiction. 
Hence it is not be possible for a pair of time-reversal conjugate singularities to annihilate each other.

\bibliographystyle{apsrev}
\bibliography{Bibliography}

\begin{thebibliography}{55}
\bibitem[{foo({\natexlab{a}})}]{footnote:continuous_phases}
\bibinfo{note}{Since the ``phases'' $\phi_n$ are treated as real-valued
  continuous functions in the repeated zone scheme, their absolute values may
  exceed $2\pi$. However, the labeling scheme guarantees that the full set of
  phase bands, in particular the Floquet bands $\epsilon_n =
  \phi(\vec{k},T)/T$, are always grouped within a window of at most $2\pi$.}
\bibitem[{foo({\natexlab{b}})}]{footnote:time_origin}
\bibinfo{note}{To see this, note that a continuous change of time origin from
  $0$ to $t'$ gives rise to a smooth deformation of the evolution operator from
  $U(t)$ to $U(t+t')U^\dagger(t')$. Given that the quasienergy spectrum is
  invariant under changes of time origin, 
  topological invariants must remain the same for any choice of time origin as
  well.}
\bibitem[{foo({\natexlab{c}})}]{footnote:zone-edge}
\bibinfo{note}{Note that the Floquet operator for a periodic evolution
  $\tilde{U}$ has only one quasienergy gap open. This gap may or may not be the
  zone-edge gap; the index of the gap is determined by continuation of the
  phase bands, see Sec.~\ref{sec:NaturalQuasiEnergyZone}.}
\bibitem[{foo({\natexlab{d}})}]{footnote:T2Singularities}
\bibinfo{note}{Note that if $\tilde U(\kv,t)$ has singularities occurring at
  $t=T/2$, they can be moved slightly away from this plane in $\kv,t$-space by
  adding a small perturbation.}
\end{thebibliography}

\end{document}